\documentclass[conference]{IEEEtran}
\IEEEoverridecommandlockouts
\usepackage{cite}
\usepackage{amsmath,amssymb,amsfonts}
\usepackage{algorithmic}
\usepackage{graphicx}
\usepackage{textcomp}
\usepackage{xcolor}
\usepackage{subfigure}
\usepackage{epsfig}
\usepackage{balance}
\usepackage[colorlinks=true, urlcolor=black, citecolor=blue, bookmarks=false]{hyperref}
\def\BibTeX{{\rm B\kern-.05em{\sc i\kern-.025em b}\kern-.08em
    T\kern-.1667em\lower.7ex\hbox{E}\kern-.125emX}} 
\usepackage{fancyhdr}

\begin{document}
\title{Hybrid Deep Embedding for Recommendations with Dynamic Aspect-Level Explanations \\
}

\newcommand{\superscript}[1]{\ensuremath{^{\textrm{#1}}}}
\def\sharedaffiliation{\end{tabular}\newline\begin{tabular}{c}}

\def \scu{\superscript{*}}
\def \uic{\superscript{\dag}}

\DeclareRobustCommand*{\IEEEauthorrefmark}[1]{%
  \raisebox{0pt}[0pt][0pt]{\textsuperscript{\footnotesize\ensuremath{#1}}}}

\author{Anonymous}
\author{\IEEEauthorblockN{Huanrui Luo\IEEEauthorrefmark{1},
Ning Yang\IEEEauthorrefmark{1}\IEEEauthorrefmark{*} \thanks{\IEEEauthorrefmark{*} Ning Yang is the corresponding author.}, 
Philip S. Yu\IEEEauthorrefmark{2}}
\IEEEauthorblockA{\IEEEauthorrefmark{1}School of Computer Science, Sichuan University,
Chengdu, China\\ Email: lolalolalola6363@gmail.com, yangning@scu.edu.cn}
\IEEEauthorblockA{\IEEEauthorrefmark{2}Department of Computer Science, University of Illinois at Chicago, Chicago, USA\\
Email: psyu@uic.edu}}
\maketitle

\thispagestyle{fancy}
\fancyhead{}
\lhead{}
\lfoot{978-1-7281-0858-2/19/\$31.00~\copyright~2019 IEEE}
\cfoot{}
\rfoot{}

\begin{abstract}
Explainable recommendation is far from being well solved partly due to three challenges. The first is the personalization of preference learning, which requires that different items/users have different contributions to the learning of user preference or item quality. The second one is dynamic explanation, which is crucial for the timeliness of recommendation explanations. The last one is the granularity of explanations. In practice, aspect-level explanations are more persuasive than item-level or user-level ones. In this paper, to address these challenges simultaneously, we propose a novel model called Hybrid Deep Embedding (HDE) for aspect-based explainable recommendations, which can make recommendations with dynamic aspect-level explanations. The main idea of HDE is to learn the dynamic embeddings of users and items for rating prediction and the dynamic latent aspect preference/quality vectors for the generation of aspect-level explanations, through fusion of the dynamic implicit feedbacks extracted from reviews and the attentive user-item interactions. Particularly, as the aspect preference/quality of users/items is learned automatically, HDE is able to capture the impact of aspects that are not mentioned in reviews of a user or an item. The extensive experiments conducted on real datasets verify the recommending performance and explainability of HDE. The source code of our work is available at \url{https://github.com/lola63/HDE-Python}.
\end{abstract}


\section{Introduction}
Explainable recommendation, which aims at making recommendations of items to users with the explanations why the items are recommended, has been attracting increasing attention of researchers due to its ability to improve the effectiveness, persuasiveness, and user satisfaction of recommender systems \cite{b9}. Although quite a few works have been proposed, explainable recommendation is still far from being well solved partly due to the following challenges:

\begin {itemize}
\item \textbf{Personalization of Preference Learning} The existing methods for explainable recommendation often assume different items have equal impact on a user preference. In practice, however, different items likely have different contributions to the learning of the preference of the same user. For example, a popular item reveals less information about the personal preference of a user than unpopular items liked by that user. Similarly, users who interact with the same item also have different contributions to the learning of the representation of that item. Therefore, we need a scheme to capture the differentiation of items/users when learning the representation (embedding) for a specific user/item.

\item \textbf{Dynamic Explanation} User preference often changes over time \cite{b15}. For example, one user might like fashions before having children, while after having children, she/he likely pays much more attention to baby products. The time-evolving preference of users suggests that to make the recommendation more proper for the occasion, a reasonable explanation for recommendations should take into consideration the dynamics of the user preference. 

\item \textbf{Aspect-Level Explanation} The existing explainable recommendation methods often generate the reason why a recommendation is made based on similarities between users or items, which leads to explanations such that "users who are similar to you like the item", or "this item is similar to the items you like" \cite{b19,b20}. In fact, finer-grained explanations are likely more convincing, for example, the aspect-level explanations such that "we recommend this movie to you because its topic matches your taste". However, it is not easy to capture aspect preference of users due to the sparsity of implicit feedbacks. The existing works often characterize the aspect preference for a specific user and aspect quality for a specific item through a counting based approach \cite{b14}, where only the aspects mentioned by reviews specific to that user or item are taken into account. In the real world, however, the aspects even though not mentioned in the user reviews not necessarily have no impact on user decision making.

\end {itemize}

In this paper, to address the above challenges simultaneously, we propose a novel model called Hybrid Deep Embedding (HDE) for aspect based explainable recommendation. The main idea of HDE is to learn the dynamic embeddings of users and items for rating prediction and the dynamic aspect preference/quality vectors for the generation of dynamic aspect-level explanations, through fusion of the dynamic implicit feedbacks extracted from reviews and the attentive user-item interactions.

First, to address the challenge of personalization of preference learning, we introduce two Personalized Embeddings (PE), to represent the personalization of users and items, respectively. PEs are learned with attention network and encode the different contributions of different items to the embedding of a user (PE of user) and the different contributions of different users to the embedding of an item (PE of item). Second, to address the challenge of dynamic explanation, we also introduce two Temporal Embeddings (TE), which are learned with LSTM\cite{b32} based network to model the sequential reviews involving a specific user (by TE of user) and those involving a specific item (by TE of item). As intermediate embeddings, PE and TE encode the personalized information and the dynamics of the preferences of users and items, respectively. By fusing the learned PEs and TEs, HDE will generate the final embeddings of users and items that are used to predict the ratings. Finally, in order to generate the dynamic aspect-level explanations for recommendations, we introduce an encoder-decoder based network by which HDE can automatically learn Aspect Preference Vectors (APV) for users and Aspect Quality Vectors (AQV) for items. APVs and AQVs can capture user preference to and item quality on aspects, respectively, even for those aspects that are not mentioned in the reviews of a specific user or item. Our main contributions are summarized as follows: 
\begin{itemize}
	\item We propose a novel model called Hybrid Deep Embedding (HDE) for aspect based explainable recommendations. By capturing dynamic personalized preferences of users to items, HDE can make recommendations with dynamic aspect-level explanations.
   \item We propose a hybrid embedding approach to learn the representations of users and items for rating prediction as well as the APVs and AQVs for dynamic aspect-level explanations.
   \item The extensive experiments on real datasets verify the recommending performance and explainability of HDE. 
\end{itemize}


\begin{flushright}
	
\end{flushright}

\section{Preliminaries and Problem Formulation}\label{section:notations and preliminaries}

\subsection{Basic Definitions}
Let $\boldsymbol{U}$ be the set of $N$ users, and $\boldsymbol{V}$ the set of $M$ items. Let $\boldsymbol{R} \in \mathbb{R}^{N \times M}$ be the rating matrix where cell at $u$-th row and $v$-th column, $R(u,v)$, represents the rating score given by user $u$ to item $v$.

We associate each user $u \in \boldsymbol{U}$ with a \textbf{user implicit feedback vector} $\boldsymbol{u} \in \{0,1\}^M$ where $v$-th component $\boldsymbol{u}(v) = 1$ if there exists an implicit feedback of $u$ to item $v$ and $\boldsymbol{u}(v) = 0$ otherwise. Here the term implicit feedback refers to user actions such as watching videos, purchasing products, and clicking items, while explicit feedback particularly refers to ratings users give to items. Similarly, we also associate each item $v \in \boldsymbol{V}$ with an \textbf{item implicit feedback vector} $\boldsymbol{v} \in \{0,1\}^N$ where $u$-th component $\boldsymbol{v}(u) = 1$ if there exists an implicit feedback to item $v$ given by user $u$ and $\boldsymbol{v}(u) = 0$ otherwise. 

For a user $\boldsymbol{u}$, we pre-train a sequence of \textbf{user review embeddings} $\langle \boldsymbol{e}^{(u,1)}, \dots, \boldsymbol{e}^{(u,T)}\rangle$, where $T$ is the maximal number of time steps considered in this paper, and $\boldsymbol{e}^{(u,t)} \in \mathbb{R}^{d_e}$ ($1 \le t \le T$) is the paragraph vector pre-trained from the review texts issued by user $u$ at time step $t$. Here $d_e$ is the dimensionality specified in advance for the pre-training of the user review vectors. We argue that the review embeddings can encode the information about the latent preference of users to the aspects of items as the reviews issued by users often contain the text mentioning the aspects. For example, the sentence "the color of this cup is nice" mentions the aspect "color" of the item cup. Similarly, for an item $v$, we also pre-train a sequence of \textbf{item review embeddings} $\langle \boldsymbol{g}^{(v,1)}, \dots, \boldsymbol{g}^{(v,T)}\rangle$, where $\boldsymbol{g}^{(v,t)} \in \mathbb{R}^{d_g}$ ($1 \le t \le T$) is the paragraph vector pre-trained from the review texts mentioning $v$ at time step $t$. And again, $d_g$ is also the dimensionality specified in advance for the pre-training of the item review embeddings. In this paper, we choose the method proposed in \cite{b4} to pre-train the review embeddings for its simplicity. However, one can note that there are many qualified paragraph embedding methods that can also serve our purpose.

As we will see later, the aspect-level explanations of recommendations will be generated based on the learned user aspect preference vectors and item aspect quality vectors. The \textbf{user aspect preference vector} of user $u$ at time step $t$ is denoted by $\boldsymbol{p}^{(u, t)} \in \mathbb{R}^f$, where $f$ is the number of aspects considered. The $i$-th component of $\boldsymbol{p}^{(u, t)}$, $\boldsymbol{p}^{(u, t)}(i)$, represents the overall preference of user $u$ to aspect $i$ at $t$. Similarly, the \textbf{item aspect quality vector} of item $v$ at time step $t$ is denoted by $\boldsymbol{q}^{(v, t)} \in \mathbb{R}^f$, where $i$-th component $\boldsymbol{q}^{(v, t)}(i)$ represents the overall preference to aspect $i$ received by item $v$ at $t$.

\begin{figure*}[t]
	\centering
	\includegraphics[width=0.95\linewidth]{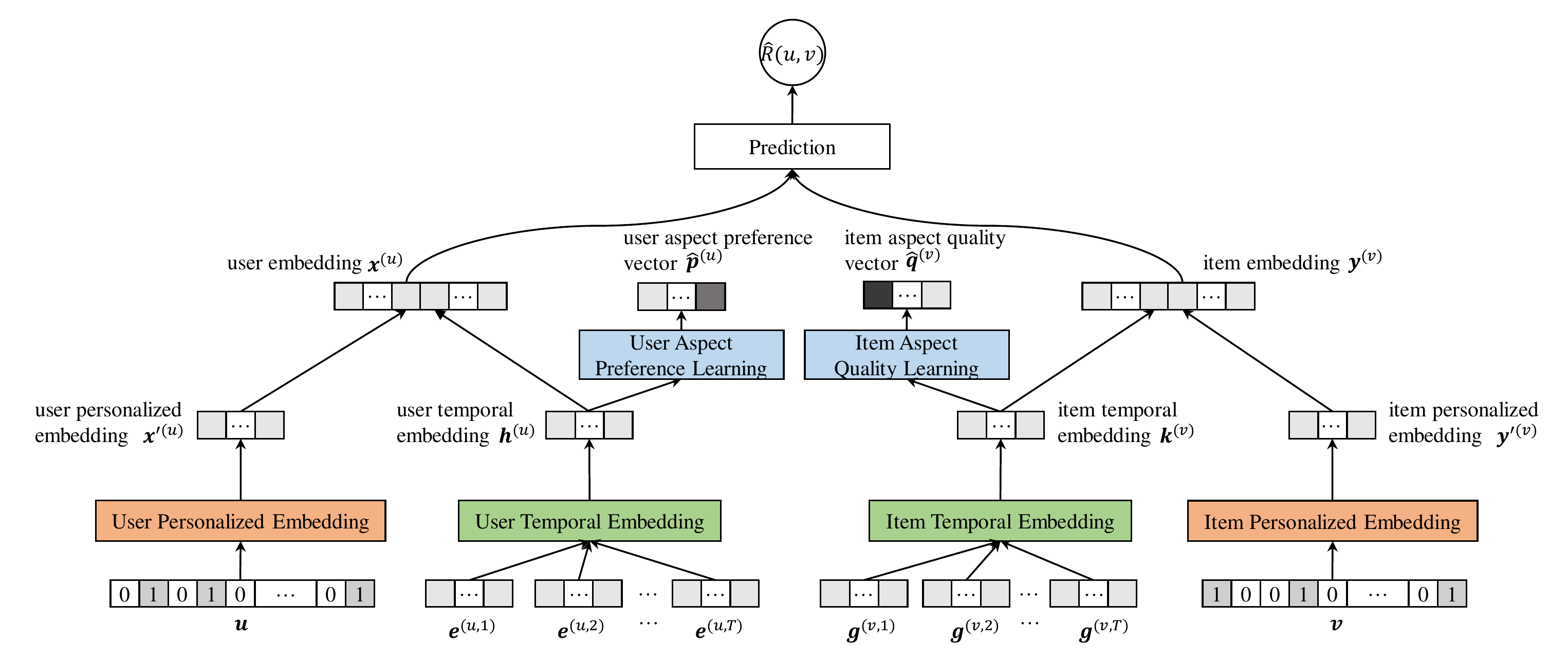}
	\caption{The architecture of HDE.}
	\label{fig:model_img}
\end{figure*}

%
%

\subsection{Problem Formulation}
Given a user implicit feedback vector $\boldsymbol{u}$, an item implicit feedback vector $\boldsymbol{v}$, the user review embeddings $\langle \boldsymbol{e}^{(u,1)},$ $ \dots, $ $\boldsymbol{e}^{(u,T)}\rangle$, and the item review embeddings $\langle \boldsymbol{g}^{(v,1)}, \dots, \boldsymbol{g}^{(v,T)}\rangle$, we want to predict the rating given by user $u$ to item $v$, $R(u, v)$, and generate the aspect preference vector $\boldsymbol{p}^{(u)} = \boldsymbol{p}^{(u, T)}$ and aspect quality vector $\boldsymbol{q}^{(v)} = \boldsymbol{q}^{(v, T)}$ for user $u$ and item $v$, based on which the aspect-level explanations can be produced for the recommendation of $v$ to $u$.

\section{Hybrid Deep Embedding}\label{section:model}

\subsection{Architecture of HDE}

The architecture of HDE is shown in Figure \ref{fig:model_img}. Given a user $u$ and item $v$, HDE will produce the prediction of the rating $\hat{R}(u, v)$ given by user $u$ to $v$ by fusing the learned user embedding $\boldsymbol{x}^{(u)}$ and item embedding $\boldsymbol{y}^{(v)}$, and at the same time, generate the explicit aspect preference vector $\boldsymbol{p}^{(u)}$ and aspect quality vector $\boldsymbol{q}^{(v)}$ used for the generation of aspect-level explanations. As we can see from Figure \ref{fig:model_img}, HDE can be roughly divided into two symmetric parts, left part and right part. The left part is responsible for learning the user embedding $\boldsymbol{x}^{(u)}$ and user aspect preference vector $\hat{\boldsymbol{p}}^{(u)}$, while the right part for learning the item embedding $\boldsymbol{y}^{(v)}$ and item aspect quality vector $\hat{\boldsymbol{q}}^{(v)}$. 

In the left part, to learn the user embedding $\boldsymbol{x}^{(u)}$, HDE first generate two intermediate embeddings for a user, one is the user personalized embedding $\boldsymbol{x}'^{(u)}$ and the other is the user temporal embedding $\boldsymbol{h}^{(u)}$. The user personalized embedding $\boldsymbol{x}'^{(u)}$ is generated by the User Personalized Embedding (UPE) component taking the user implicit feedback vector $\boldsymbol{u}$ as input. As we will see later, thanks to the attention network in the UPE, the generated user personalized embedding $\boldsymbol{x}'^{(u)}$ can capture the different contributions of different items to a specific user, which is crucial for the personalization of preference learning of HDE. At the same time, HDE will generate the user temporal embedding $\boldsymbol{h}^{(u)}$ through the User Temporal Embedding (UTE) component. UTE is an LSTM-based network with the sequence of the pre-trained user review embeddings $\langle \boldsymbol{e}^{(u,1)},$ $ \dots, $ $\boldsymbol{e}^{(u,T)}\rangle$ as input. Here we can regard UTE as an encoder which encodes the dynamic aspect information from the reviews into the user temporal embedding $\boldsymbol{h}^{(u)}$. At last, HDE will generate the user embedding $\boldsymbol{x}^{(u)}$ by fusing the two intermediate embeddings $\boldsymbol{x}'^{(u)}$ and $\boldsymbol{h}^{(u)}$. We argue that $\boldsymbol{x}^{(u)}$ encodes not only the information about the user personalized preference but also the information about the dynamics of the user preference. One can also note that the user temporal embedding $\boldsymbol{h}^{(u)}$ is also fed into the User Aspect Preference Learning (UAPL) component, which is a fully connected network and can be regarded as a decoder corresponding to UTE, to produce the explicit user aspect preference vector $\hat{\boldsymbol{p}}^{(u)}$ where each dimension represents an aspect.

Symmetrically, in the right part, for an item $v$, HDE also generates two intermediate embeddings, the item personalized embedding $\boldsymbol{y}'^{(v)}$ and the item temporal embedding $\boldsymbol{k}^{(v)}$, through the Item Personalized Embedding (IPE) component and the Item Temporal Embedding (ITE) component, respectively, and then produce the item embedding $\boldsymbol{y}^{(v)}$ by fusing them. At the same time, HDE also generate the explicit aspect quality vector $\hat{\boldsymbol{q}}^{(v)}$ through the Item Aspect Quality Learning (IAQL) component. Note that IPE, ITE, and IAQL are the counterparts of UPE, UTE, and UAPL, except that the input of ITE is the item implicit feedback vector $\boldsymbol{v}$ and the input of ITE is the item review embeddings $\langle \boldsymbol{g}^{(v,1)}, \dots, \boldsymbol{g}^{(v,T)}\rangle$.

\subsection{Personalized Embedding}\label{section:Personlized Embedding Layer}
\begin{figure}[h]
	\centering
	\includegraphics[width=0.8\linewidth]{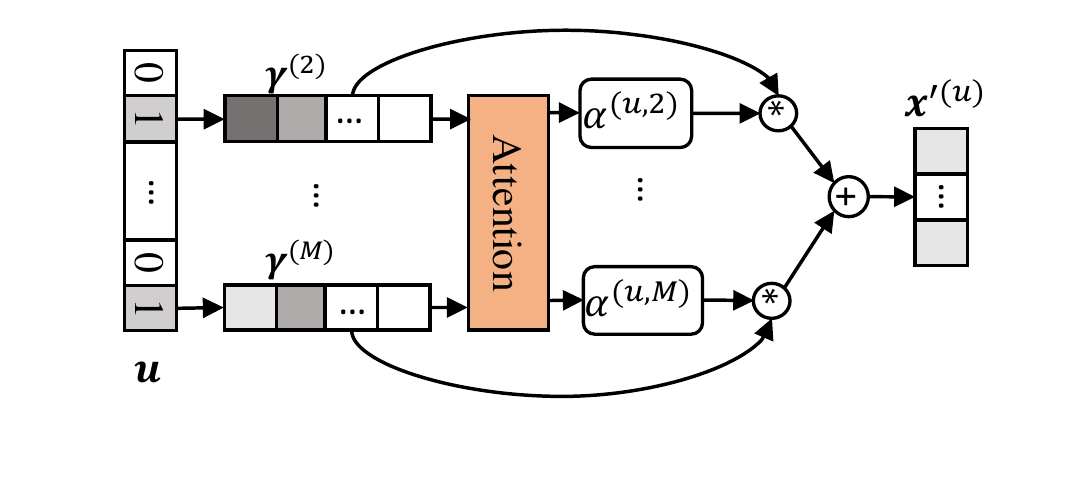}
	\caption{Personalized Embedding.}
	\label{fig:personlized_embedding_img}
\end{figure}
The goal of UPE and IPE is to capture the personalized preference of users offered to items, and the personalized preference of items received from users, to generate the personalized embeddings $\boldsymbol{x}'^{(u)} \in \mathbb{R}^{d_p}$ and $\boldsymbol{y}'^{(v)} \in \mathbb{R}^{d_p}$, respectively, where $d_p$ is the dimensionality of the personalized embedding. Due to the symmetry, here we just describe UPE in detail and IPE has the similar structure.

Intuitively, the personalized preference of a user to items is indicated by her/his interactions with items, which are represented by the user implicit feedback vector $\boldsymbol{u}$. Let $J(u)$ be the set of items interacted with user $u$. Then the $v$-th component $\boldsymbol{u}(v) = 1$ if $v \in J(u)$, otherwise $\boldsymbol{u}(v) = 0$. However, it is reasonable that different items have different contributions to the user personalized preference. To capture such difference, we introduce an attentional network to the UPE, whose structure is shown in Figure \ref{fig:personlized_embedding_img}. For each item $v \in J(u)$, HDE represents it with an item latent vector $\boldsymbol{\gamma}^{(v)} \in \mathbb{R}^{d_{\gamma}}$, where $d_{\gamma}$ is the dimensionality. According to Figure \ref{fig:personlized_embedding_img}, the user personalized embedding $\boldsymbol{x}'^{(u)}$ is calculated as:
\begin{equation}
\boldsymbol{x}'^{(u)} = \sum_{v\in{J(u)}}\alpha^{(u,v)}\boldsymbol{\gamma}^{(v)}
\label{Eq_UPE},
\end{equation}
where $\alpha^{(u,v)}$ is the attention score, which can be interpreted as the contribution of item $v$ to user $u$. The attention score is calculated as follows:
\begin{equation}
\alpha^{(u,v)}=\dfrac{\exp{(s'^{(v)})}}{\sum_{v\in{J(u)}}\exp{(s'^{(v)})}},
\label{Eq_Attention}
\end{equation}
\begin{equation}
{s}'^{(v)} = \boldsymbol{s}^T\tanh(\boldsymbol{W}_{s}\boldsymbol{\gamma}^{(v)}+\boldsymbol{b}_{s}),
\label{Eq_Sv}
\end{equation}
where $\boldsymbol{s} \in \mathbb{R}^{d_a}$ is the query vector of dimensionality $d_a$ specified in advance. Note that in Equations (\ref{Eq_UPE}), (\ref{Eq_Attention}), and (\ref{Eq_Sv}), $\boldsymbol{\gamma}^{(v)}$, $\boldsymbol{s} \in \mathbb{R}^{d_a}$, $\boldsymbol{W}_{s }\in \mathbb{R}^{{d_a} \times d_\gamma}$, and $\boldsymbol{b}_{s}\in\mathbb{R}^{d_a}$ will be learned during the model learning.

Symmetrically, IPE has the similar structure with UPE. Let $Q(v)$ be the set of users who have interacted with item $v$. Then the $u$-th component of the item implicit feedback vector $\boldsymbol{v}$, $\boldsymbol{v}(u) = 1$ if $u \in Q(v)$, otherwise $\boldsymbol{v}(u) = 0$.  For a user $u$, HDE also represents it with a user latent vector $\boldsymbol{\mu}^{(u)} \in \mathbb{R}^{d_{\gamma}}$. Then the personalized embedding of item $v$, $\boldsymbol{y}'^{(v)}$, can be calculated as:
\begin{equation}
\boldsymbol{y}'^{(v)} = \sum_{u\in{Q(v)}}\beta^{(v,u)}\boldsymbol{\mu}^{(u)},
\label{Eq_IPE}
\end{equation}
where $\beta^{(v,u)}$ is the attention score of user $u$ to item $v$. Similarly, $\beta^{(v,u)}$ can be obtained through the following equations which are similar to Equations (\ref{Eq_Attention}) and (\ref{Eq_Sv}):
\begin{equation}
\beta^{(v,u)}=\dfrac{\exp{(r'^{(u)})}}{\sum_{u\in{Q(v)}}\exp{(r'^{(u)})}},
\label{Eq_Attention_Item}
\end{equation}
\begin{equation}
{r}'^{(u)} = \boldsymbol{r}^T\tanh(\boldsymbol{W}_{r}\boldsymbol{\mu}^{(u)}+\boldsymbol{b}_{r}),
\label{Eq_Rv}
\end{equation}
where $\boldsymbol{r} \in \mathbb{R}^{d_a}$ is the query vector. Similarly, in Equations (\ref{Eq_IPE}), (\ref{Eq_Attention_Item}), and (\ref{Eq_Rv}), $\boldsymbol{\mu}^{(u)}$, $\boldsymbol{r}$, $\boldsymbol{W}_{r} \in \mathbb{R}^{{d_a} \times d_r}$, and $\boldsymbol{b}_{r}\in\mathbb{R}^{d_a}$ will also be learned during the model learning.

\begin{figure}[t]
	\centering
	\includegraphics[width=0.95\linewidth]{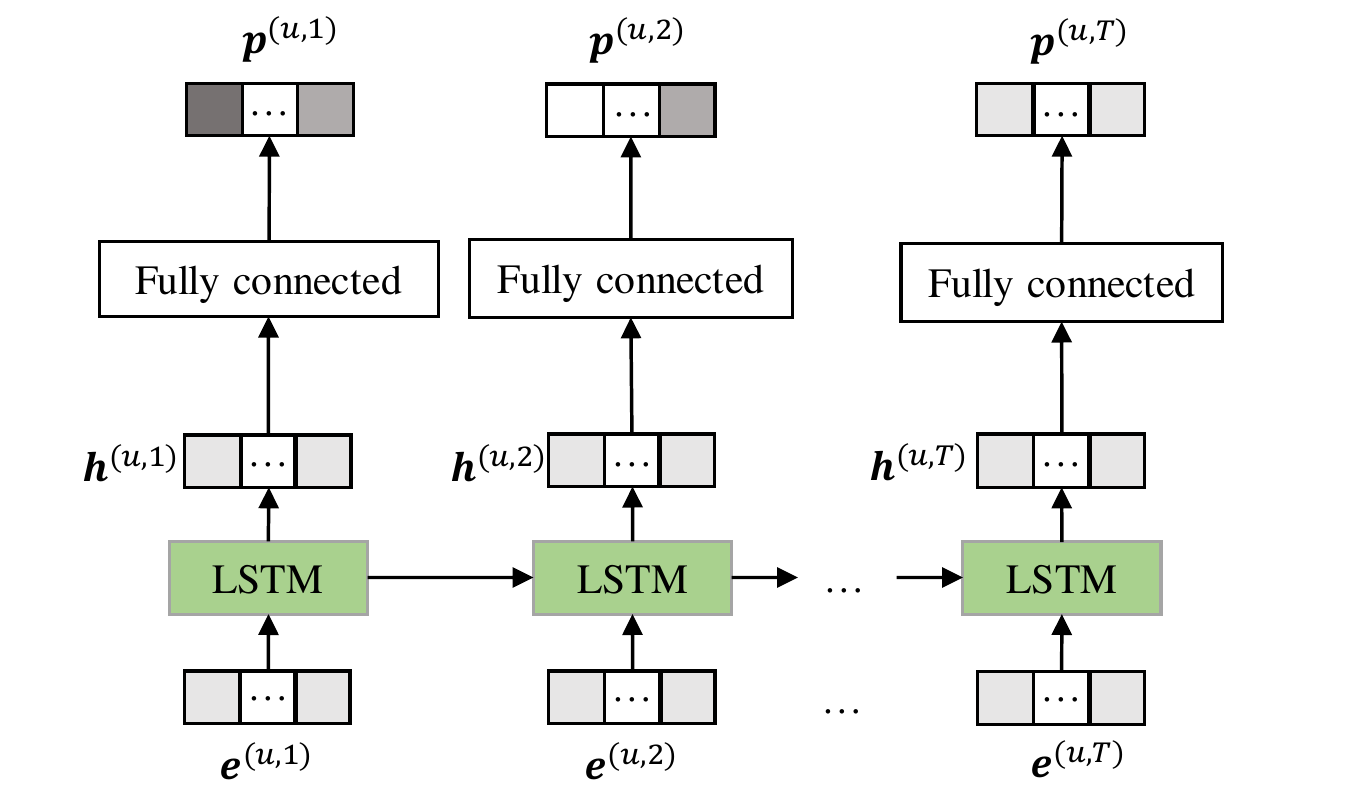}
	\caption{Temporal Embedding and Aspect Preference Learning.}
	\label{fig:temporal_embedding_img}
\end{figure}

\begin{figure}[t]
	\centering
	\includegraphics[width=0.8\linewidth]{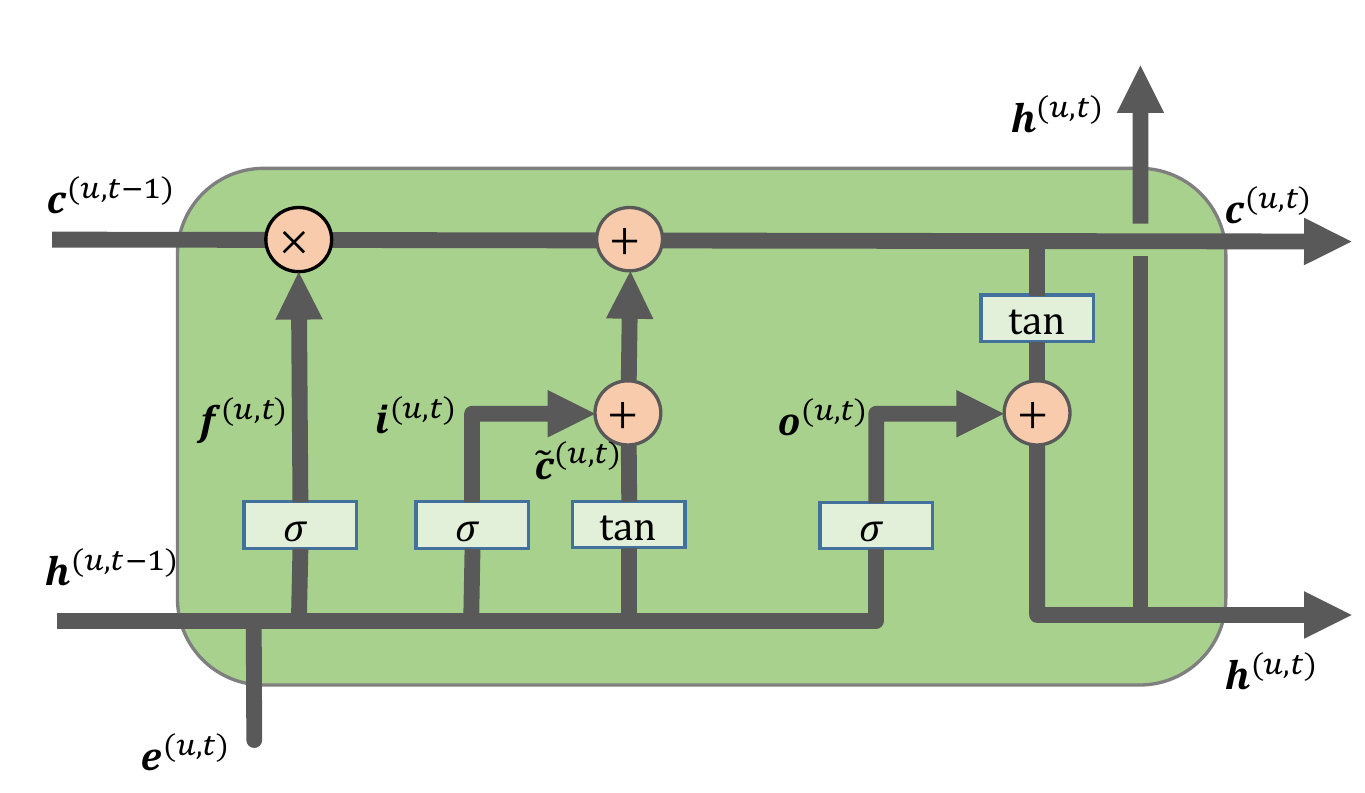}
	\caption{LSTM.}
	\label{fig:lstm_img}
\end{figure}

\subsection{Temporal Embedding}\label{section:Temporal Embedding layer}
As shown in the bottom part of Figure \ref{fig:temporal_embedding_img}, UTE is an LSTM based network, by which the dynamic aspect-level preference hidden in the sequence of review embeddings ($\boldsymbol{e}^{(u,t)}$, $1 \le t \le T$) can be encoded into the user temporal embedding $\boldsymbol{h}^{(u, t)} \in \mathbb{R}^{d_t}$ for some user $u$, where $d_t$ is the dimensionality of temporal embedding. Again due to the symmetry, with the similar structure ITE can take the sequence of review embeddings ($\boldsymbol{g}^{(v,t)}$, $1 \le t \le T$) and generate the item temporal embedding $\boldsymbol{k}^{(v, t)} \in \mathbb{R}^{d_t}$ for some item $v$.

Figure \ref{fig:lstm_img} shows the detail of an LSTM unit by which a user temporal embedding $\boldsymbol{h}^{(u, t)}$ can be produced via the following equations:
\begin{equation}
\label{Eq_LSTM}
\begin{split}
\boldsymbol{f}^{(u,t)} &=\sigma(\boldsymbol{W}_{f}\cdot[\boldsymbol{h}^{(u,t-1)},\boldsymbol{e}^{(u,t)}]+\boldsymbol{b}_{f})\\
\boldsymbol{i}^{(u,t)} &=\sigma(\boldsymbol{W}_{i}\cdot[\boldsymbol{h}^{(u,t-1)},\boldsymbol{e}^{(u,t)}]+\boldsymbol{b}_{i})\\
\boldsymbol{\tilde{c}}^{(u,t)} &= \tanh(\boldsymbol{W}_{c}\cdot[\boldsymbol{h}^{(u,t-1)},\boldsymbol{e}^{(u,t)}]+\boldsymbol{b}_{c})\\
\boldsymbol{c}^{(u,t)} &=\boldsymbol{f}^{(u,t)}*\boldsymbol{c}^{(u,t-1)}+\boldsymbol{i}^{(u,t)}*\boldsymbol{\tilde{c}}^{(u,t)}\\
\boldsymbol{o}^{(u,t)} &=\sigma(\boldsymbol{W}_{o}\cdot[\boldsymbol{h}^{(u,t-1)},\boldsymbol{e}^{(u,t)}]+\boldsymbol{b}_{o})\\
\boldsymbol{h}^{(u,t)} &=\boldsymbol{o}^{(u,t)}*\tanh\boldsymbol{c}^{(u,t)},
\end{split}
\end{equation}
where $\boldsymbol{f}^{(u,t)}$, $\boldsymbol{i}^{(u,t)}$, and $\boldsymbol{o}^{(u,t)}$ denote forget gate, input gate, and output gate, respectively, and $\boldsymbol{c}^{(u,t)}$ is the cell activation vector. $\boldsymbol{W}_{f}$, $\boldsymbol{W}_{i}$, $\boldsymbol{W}_{c}$, $\boldsymbol{W}_{o} \in\mathbb{R}^{{d_t}\times(d_t+d_e)} $, $\boldsymbol{b}_{f}$, and $\boldsymbol{b}_{i}$, $\boldsymbol{b}_{c}$, $\boldsymbol{b}_{o} \in\mathbb{R}^{d_t} $ are the parameters that will be learned during the model training. Note that ITE has the same structure as UTE except that ITE has its own parameters and takes sequence of review embeddings $\boldsymbol{g}^{(v,t)}$ as input.

\subsection{Aspect Preference/Quality Learning}
As the user temporal embeddings $\boldsymbol{h}^{(u,t)}$ and item temporal embeddings $\boldsymbol{k}^{(v,t)}$ carry the latent dynamic aspect-level preference hidden in reviews, they will be fed into their respective decoders, the user aspect preference learning (UAPL) and the item aspect quality learning (IAQL), to produce the explicit user aspect preference vector $\boldsymbol{\hat{p}}^{(u,t)}\in\mathbb{R}^{f}$ and item aspect quality vector $\boldsymbol{\hat{q}}^{(u,t)}\in\mathbb{R}^{f}$, respectively, where $f$ is the number of aspects. The user aspect preference vectors and the item aspect quality vectors will be further used to generate the aspect-level explanations for a recommendation.

UAPL and IAQL are both a fully connected network, which can generate the user aspect preference vector $\boldsymbol{\hat{p}}^{(u,t)} \in \mathbb{R}^f$ for a user $u$ and item aspect quality vector $\boldsymbol{\hat{q}}^{(v,t)}  \in \mathbb{R}^f$ for an item $v$ respectively using the equation
\begin{equation}
\boldsymbol{\hat{p}}^{(u,t)}=\sigma(\boldsymbol{W}_{p}\cdot\boldsymbol{h}^{(u,t)}+\boldsymbol{b}_{p})
\label{Eq_p}
\end{equation}
and equation
\begin{equation}
\boldsymbol{\hat{q}}^{(v,t)}=\sigma(\boldsymbol{W}_{q}\cdot\boldsymbol{k}^{(v,t)}+\boldsymbol{b}_{q}),
\label{Eq_q}
\end{equation}
where $\boldsymbol{W}_{p}$, $\boldsymbol{W}_{q}\in\mathbb{R}^{{f} \times d_t}$ and $\boldsymbol{b}_{p}$, $\boldsymbol{b}_{q}\in\mathbb{R}^{f}$ are the parameters to be learned.

\subsection{Rating Prediction}
Now we have produced two intermediate embeddings, personalized embedding and temporal embedding, for a user $u$ and an item $v$. The personalized embeddings $\boldsymbol{x}'^{(u)}$ and $\boldsymbol{y}'^{(v)}$ capture the attentional personalized preference of user $u$ giving to different items and the attentional personalized preference of item $v$ receiving from different users, respectively, while the temporal embeddings $\boldsymbol{h}^{(u)} = \boldsymbol{h}^{(u, T)}$ and $\boldsymbol{k}^{(v)} = \boldsymbol{k}^{(v, T)}$ encode the dynamic aspect preference information of user $u$ and item $v$, respectively. 

In order to fuse the personalized preference and the dynamic aspect preference simultaneously, HDE will generate the final user embedding $\boldsymbol{x}^{(u)}$ and the final item embedding $\boldsymbol{y}^{(v)}$ for a user $u$ and an item $v$ with the following equations, respectively,
\begin{eqnarray} 
\boldsymbol{x}^{(u)} &= \boldsymbol{x}'^{(u)}\oplus\boldsymbol{h}^{(u)} , \text{ }
\boldsymbol{y}^{(v)} &= \boldsymbol{y}'^{(v)}\oplus\boldsymbol{k}^{(v)}, 
\end{eqnarray}
where $\oplus$ is concatenation operator, and $\boldsymbol{x}^{(u)}$, $\boldsymbol{y}^{(v)} \in\mathbb{R}^{d_p+d_t}$. Finally, HDE will predict the rating of user $u$ to item $v$, $\hat{R}(u,v)$, via a simple Neural Collaborative Filtering (NCF) model, i.e., 
\begin{equation}
\hat{R}_{u,v}=\phi(\boldsymbol{W}_{\phi} (\boldsymbol{x}^{(u)} \odot \boldsymbol{y}^{(v)}) + \boldsymbol{b}_\phi),
\end{equation}
where $\odot$ represents the element-wise product of vectors, $\phi(\cdot)$ is the ReLU function, and $\boldsymbol{W}_{\phi} \in \mathbb{R}^{d_\phi \times (d_p + d_t)}$, $\boldsymbol{b}_\phi \in \mathbb{R}^{d_p + d_t}$ are the parameters to be learned. 

\begin{table*}[t]
	\caption{Statistics of the datasets.}
	\label{lable:dataset}
	\centering
	\begin{tabular}{llllllllll|}
		\hline
		Datasets & \#Users & \#Items & \#Reviews & \#Aspects ($f$) & \#Density \\
		\hline
		Digital Music & 5,541&3,568&64,706&98&0.33\%\\
		Video Game & 24,303&10,672&231,780&57&0.09\%\\
		Movie & 123,960&50,052&1,679,533&120&0.03\%\\
		\hline
	\end{tabular}
\end{table*}

\begin{table*}[t]
	\caption{Comparison of the Baselines.}
	\label{table:compare}
	\centering
	\begin{tabular}{l l l l l l l l l l}
		\hline
		Characteristics&PMF&HFT&EFM&DeepCoNN&NARRE&AMF&HDE&NA-HDE&NL-HDE\\
		\hline
		Ratings & $\surd$ & $\surd$ & $\surd$ & $\surd$ & $\surd$ &$\surd$ & $\surd$ & $\surd$ & $\surd$ \\
		Textual Reviews & $\backslash$ & $\surd$ & $\surd$ & $\surd$ &$\surd$ & $\surd$ & $\surd$ &$\surd$ &$\surd$ \\
		Deep Learning & $\backslash$ & $\backslash$ & $\backslash$ & $\surd$ & $\surd$ & $\surd$ & $\surd$ & $\surd$ & $\surd$\\
		Explainable & $\backslash$ & $\backslash$ & $\surd$ & $\backslash$ & $\surd$ & $\surd$ & $\surd$ & $\surd$ & $\surd$\\
		Temporal features & $\backslash$ & $\backslash$ & $\backslash$ & $\backslash$ & $\backslash$ &$\backslash$ & $\surd$ & $\surd$ & $\backslash$ \\
		\hline
	\end{tabular}
\end{table*}

\subsection{Model Training}
Let $\mathcal{I}_{train}$ be the training set consisting of user-item pairs $(u, v)$ where $u \in \boldsymbol{U} $ and $v \in \boldsymbol{V}$. Then the loss function for HDE learning is 
\begin{equation}
\begin{split}
\mathcal{L} = &\sum_{(u,v)\in{\mathcal{I}}_{train}} \big\{(\hat{R}(u,v)-R(u,v))^2\\
&+\lambda_1\sum_{t=1}^{T}[(\boldsymbol{\hat{p}}^{(u,t)}-\boldsymbol{{p}}^{(u,t)})^2+(\boldsymbol{\hat{q}}^{(v,t)}-\boldsymbol{{q}}^{(v,t)})^2] \big\}\\
&+\lambda_2 \mathcal{L}_{reg}.
\label{eq:objective}
\end{split}
\end{equation}
where $R(u,v)$ is the ground-truth of the rating, and $\lambda_1$ and $\lambda_2$ are hyper-parameters that regulate the contribution of different terms to the loss. $\mathcal{L}_{reg}$ is the regularization term which uses $L^2$-norm for all parameters to avoid overfitting. 
 
In Equation (\ref{eq:objective}), $\boldsymbol{{p}}^{(u,t)}$ and $\boldsymbol{{q}}^{(v,t)}$ are the supervisions of user explicit aspect preference vectors and item explicit aspect quality vectors, respectively, which are obtained with the method proposed by \cite{b14}. Particularly, the preference to aspect $i$ of user $u$ at time $t$, $\boldsymbol{{p}}^{(u,t)}(i)$, is computed with the following equation \cite{b14}:

\begin{equation}
\label{Eq_True_P}
\boldsymbol{p}^{(u,t)}(i)=
\begin{cases}
0,\ \text{if user } u \text{ does not mention aspect } i \\ \text{\ \ \ \ until time } t \\
1+(a-1)(\dfrac{2}{1+e^{-n_{i}^{(u,t)}}}), \text{otherwise},
\end{cases}
\end{equation}
where $a$ is the maximum value that a rating can be (usually $a = 5$), and $n_{i}^{(u,t)}$ is the total number of times that user $u$ mentions aspect $i$ till $t$. The idea here is that the more frequently (i.e., larger $n_{i}^{(u,t)}$) the aspect $i$ is mentioned by $u$, the greater the preference of $u$ to aspect $i$. Similarly, the quality of aspect $i$ of item $v$ at time $t$, $\boldsymbol{{q}}^{(v,t)}(i)$, is computed with the following equation \cite{b14}:
\begin{equation}
\label{Eq_True_Q}
\boldsymbol{q}^{(v,t)}(i)=
\begin{cases}
0,\ \text{if aspect } i \text{ of item } v  \text{ is not mentioned} \\ \text{\ \ \ \ until time } t \\
1+\dfrac{a-1}{1+e^{-{k}_{i}^{(v,t)}\cdot{s_{i}^{(v,t)}}}}, \text{otherwise},
\end{cases}
\end{equation}
where $k_{i}^{(v,t)}$ is the total number of times that aspect $i$ of item $v$ is mentioned till time $t$, and $s_{i}^{(v,t)}$ represents the average sentiment of the reviews on aspect $i$ of item $v$ till time $t$. We use the following equation to produce $s_{i}^{(v,t)}$ \cite{b14}:

\begin{equation}
s_{i}^{(v,t)}=\dfrac{\sum_{j}^{{k}_{i}^{(v,t)}} {S}_{j}^{(v,i)}}{{k}_{i}^{(v,t)}}
\end{equation} 
where $S_{j}^{(v,i)}$ is 1 if the aspect $i$ of item $v$ is mentioned with positive opinion words at $j$-th review, and -1 otherwise. Before the training of HDE, the opinion words and aspect words will be extracted with the method used in \cite{b14}. As the extraction of the words is not the focus of this paper, we refer the interested readers to \cite{b14} for more details.

\section{Evaluation of Rating Prediction}\label{section:experiments}
\subsection{Experimental Setting}
\subsubsection{Datasets}
The experiments are conducted on three real-world datasets collected from Amazon, Digital Music, Video Game, and Movie, all of which contain user-item ratings and textual reviews. The statistics of the datasets are presented in Table \ref{lable:dataset}. The aspects on the datasets are extracted with the same method as used in \cite{b14}, which generates the aspect words from the text review corpus using grammatical and morphological analysis tools. Particularly, the number of aspects $f = 98$, 57, and 120 on Digital Music, Video Game, and Movie, respectively. On each dataset, we randomly select 80\% as training set, 10\% as validation set, and the remaining 10\% as testing set.
\renewcommand\arraystretch{2}

\begin{figure}[t]
	\centering
	\subfigure[RMSE.]{
		\includegraphics[width=0.46\linewidth]{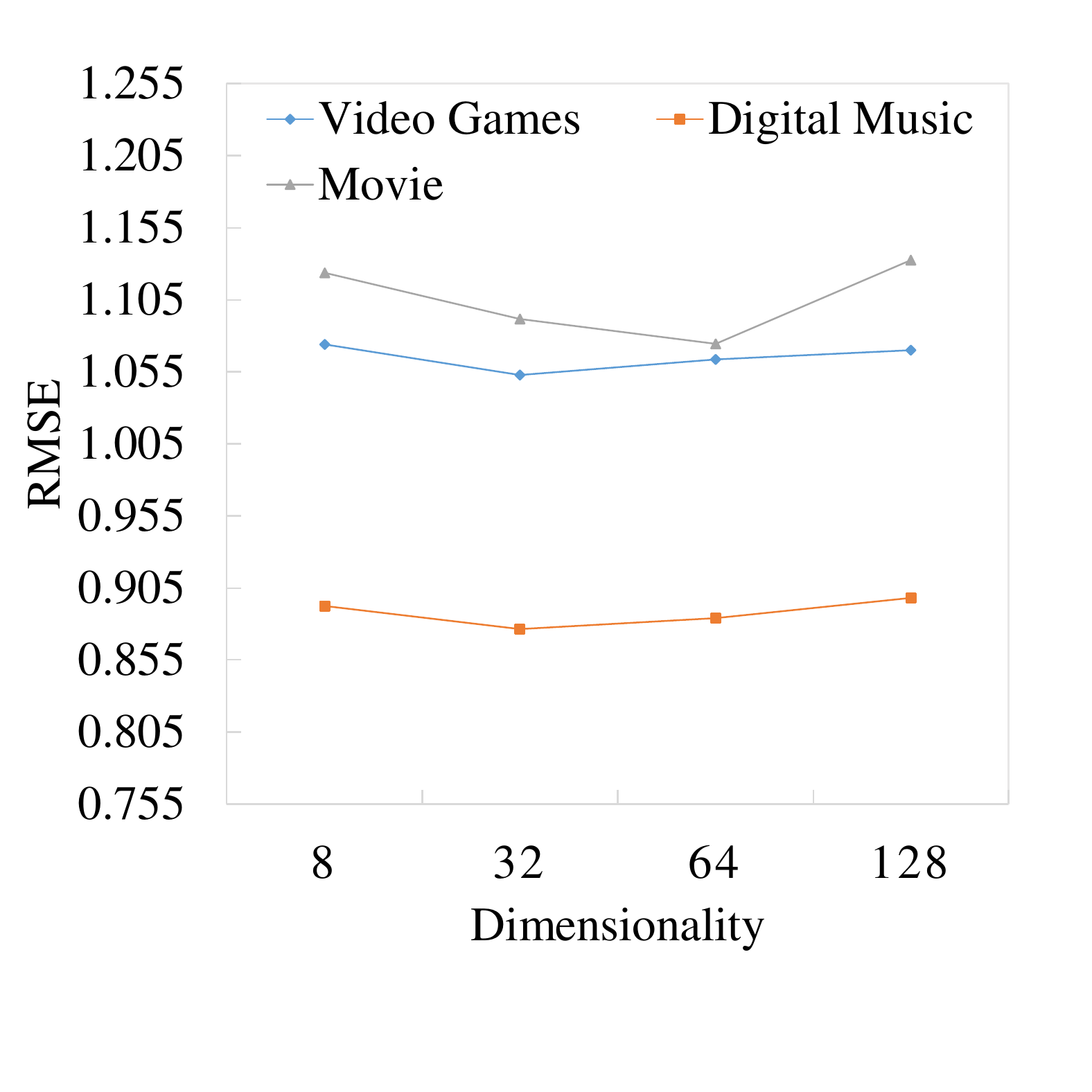}
	}
	\subfigure[MAE.]{
		\includegraphics[width=0.46\linewidth]{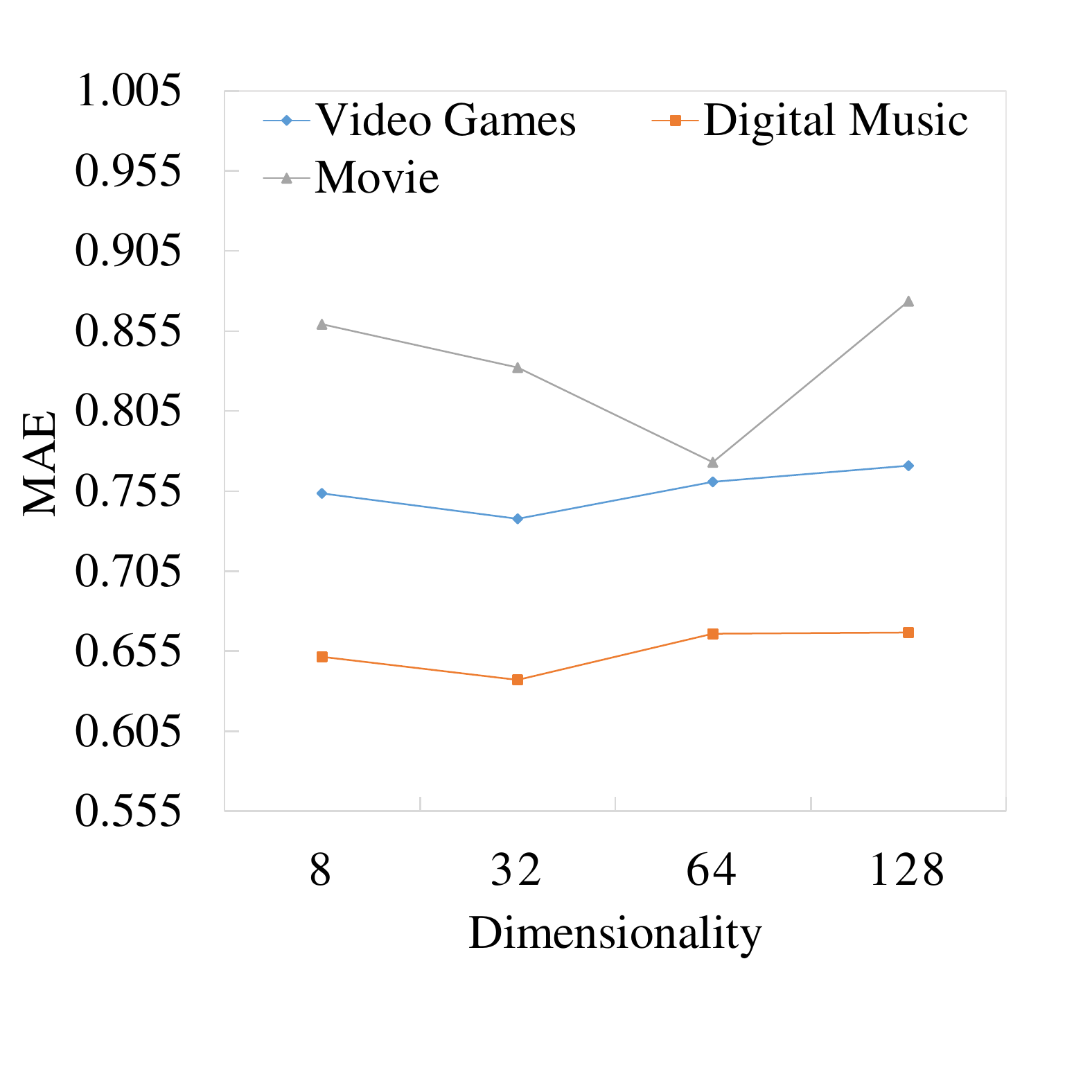}
	}
	\caption{Tuning of Embedding Dimensionality}
	\label{Sensitivity_factor:Digit}
\end{figure}

\subsubsection{Baselines}
In order to demonstrate the effectiveness of HDE, we compare our model with the following five models, PMF, HFT, EFM, DeepCoNN, NARRE, and AMF, whose characteristics are showed in Table \ref{table:compare}. \par
\renewcommand\arraystretch{2}

\begin{itemize}
	\item \textbf{PMF}\cite{b7}  Probabilistic Matrix Factorization (PMF) is a classic factor based recommendation algorithm which models the user preference matrix as a product of two lower-rank user and movie matrices.
	
	\item \textbf{HFT} \cite{b6} Hidden Factors and hidden Topics (HFT) model can make product recommendations with a fusion of ratings and review texts. Particularly, HFT uses LDA\cite{b8} method to obtain the stochastic topic distribution of reviews, and combines it with a latent factor model.
	
	\item \textbf{EFM}\cite{b14} Explicit Factor Model (EFM) is an explainable recommendation model which first extracts aspects and user opinions by phrase-level sentiment analysis on user reviews, and then generates with aspect-level explanations.
	
	\item \textbf{DeepCoNN}\cite{b10}  DeepCoNN utilizes two parallel CNN networks to process reviews, one for the modeling of user's behavioral features, and the other for the reviews received by the item, and jointly models users and items by a Factorization Model.
	
	\item \textbf{NARRE}\cite{b1} NARRE is a neural attentional regression model with review-level explanations (NARRE) for recommendation, which introduces an attention mechanism to explore the usefulness of reviews.
	
	\item \textbf{AMF}\cite{b3}  AMF is an aspect-based latent factor model which can make recommendations by fusing explicit feedbacks of users with auxiliary aspect information extracted from reviews of items. 
	
\end{itemize}

Additionally, in order to verify the effectiveness of the Personalized Embedding component and the Temporal Embedding component of HDE, we also compare HDE with two more baseline methods, NA-HDE and NL-HDE. NA-HDE is a variant of HDE removing the personalized embedding component, while NL-HDE is a variant of HDE removing the temporal embedding component.

\subsubsection{Evaluation Metrics} 
We use the Root Mean Square Error (RMSE) and Mean Absolute Error (MAE) as the evaluation metrics, which are defined as:
\begin{equation}
RMSE=\sqrt{\frac{\sum_{(u,v) \in \mathcal{I}_{test}}(R(u,v)-\hat{R}(u,v))^2}{|\mathcal{I}_{test}|}},
\end{equation}
\begin{equation}
MAE=\frac{\sum_{(u,v) \in \mathcal{I}_{test}}|R(u,v)-\hat{R}(u,v)|}{|\mathcal{I}_{test}|},
\end{equation} 
where $\mathcal{I}_{test}$ is the testing set.

\subsubsection{Parameter Setting}
The hyper-parameters are tuned on the validation set. We set the batch size as $128$, the dropout ratio 0.3. For simplicity, we set the dimensionalities $d_\phi$, $d_a$, $ d_e$, $d_p$, $d_t$, and $d_{\gamma}$ with the same value on the same dataset. Figure \ref{Sensitivity_factor:Digit} shows that both RMSE and MAE achieve the best at the dimensionality of 32 on Digit Music and Video Game, while 64 on Movie. Therefore we set $d_\phi$$=d_a$$ = d_e$$=d_p$$=d_t$$ = d_{\gamma}$$ = 32$ on both Digital Music and Video Games, while 64 on Movie. However, note that theoretically the dimensionality of different embedding can be set to different value.

\begin{figure*}[t]
	\centering
	\subfigure[Digit Music.]{
		\includegraphics[width=0.25\linewidth]{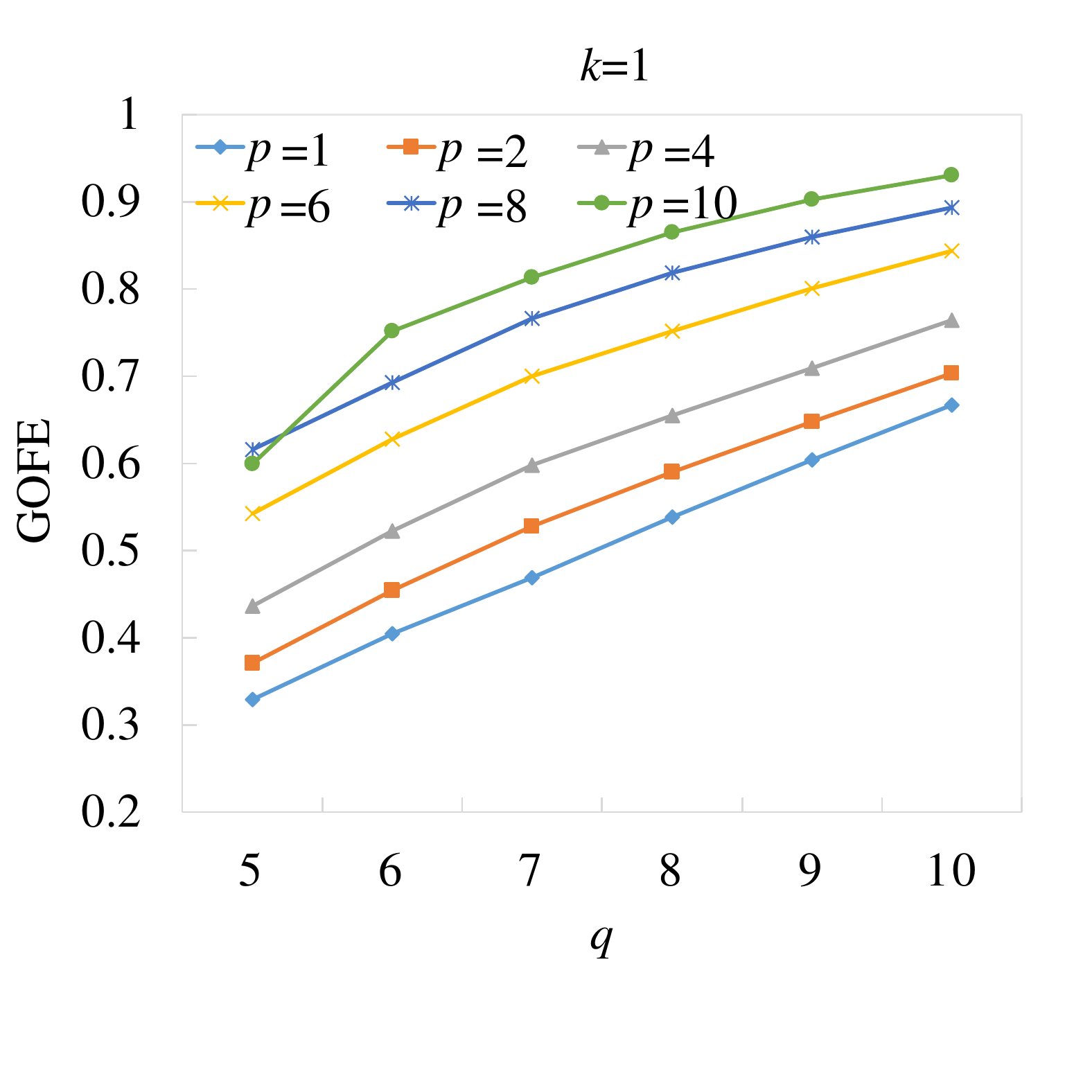}
	}
	\subfigure[Video Game.]{
		\includegraphics[width=0.25\linewidth]{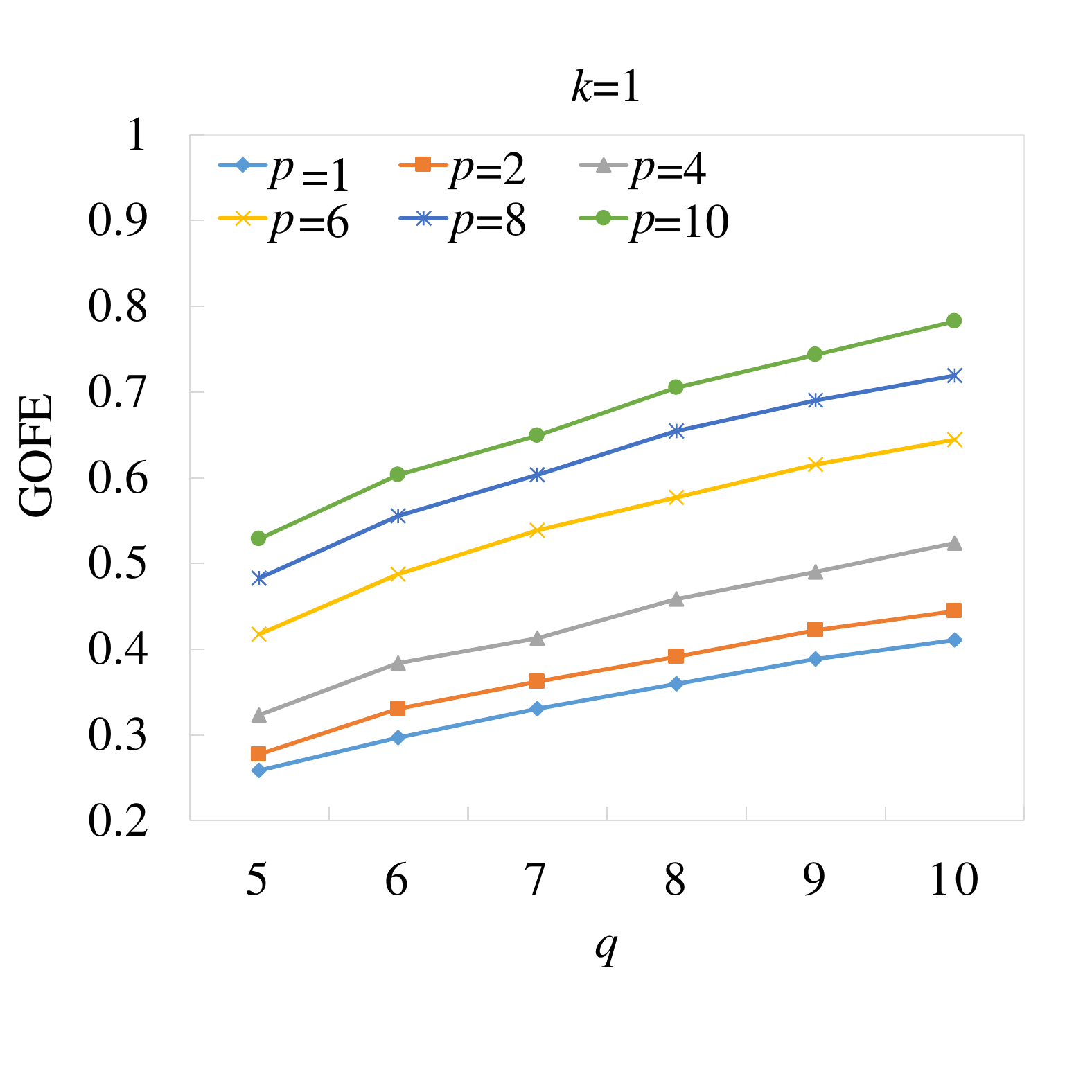}
	}
	\subfigure[Movie.]{
		\includegraphics[width=0.25\linewidth]{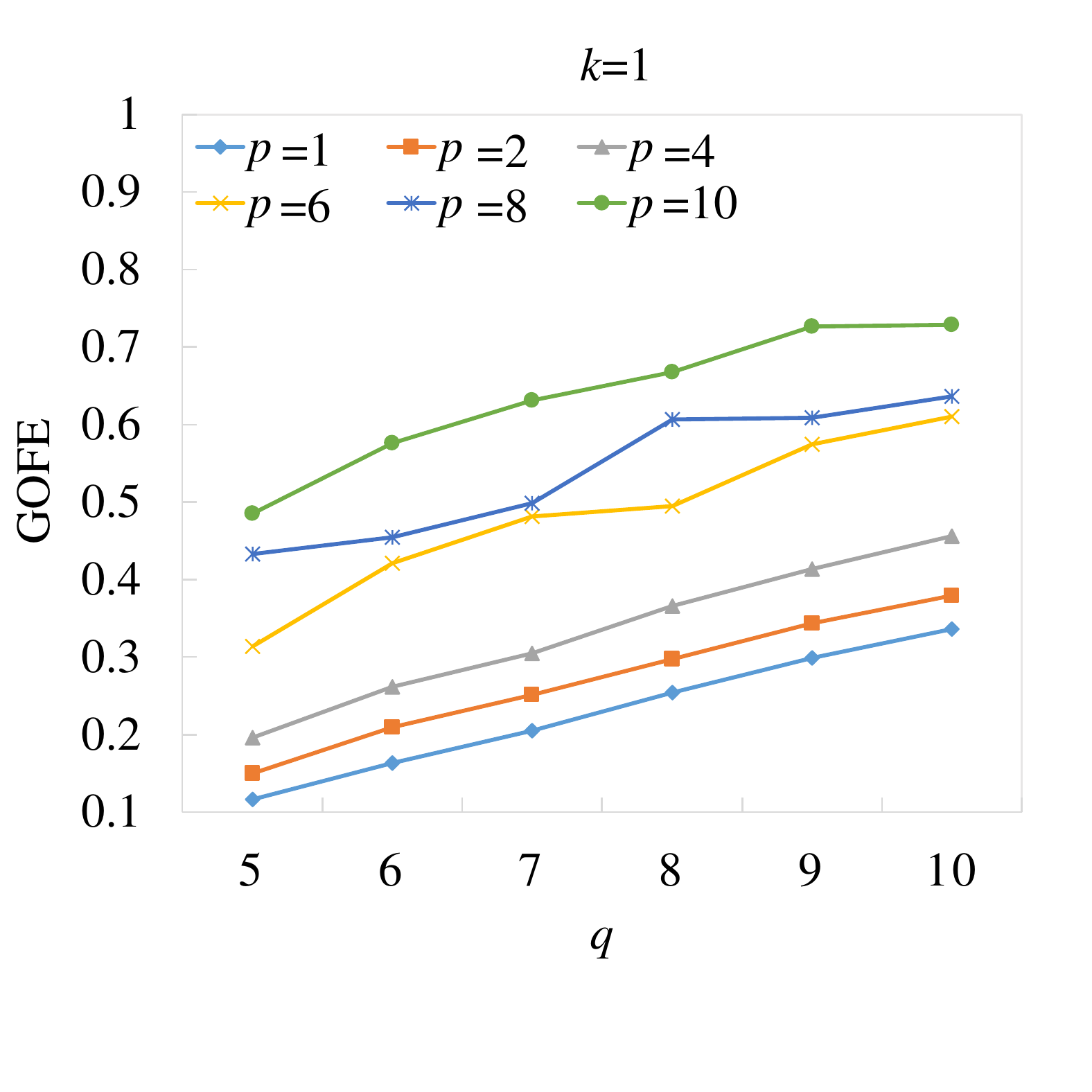}
	}
	\caption{ GOFE at $k=1$.}
	\label{explain:K=1}
\end{figure*}

\begin{figure*}[t]
	\centering
	\subfigure[Digit Music.]{
		\includegraphics[width=0.25\linewidth]{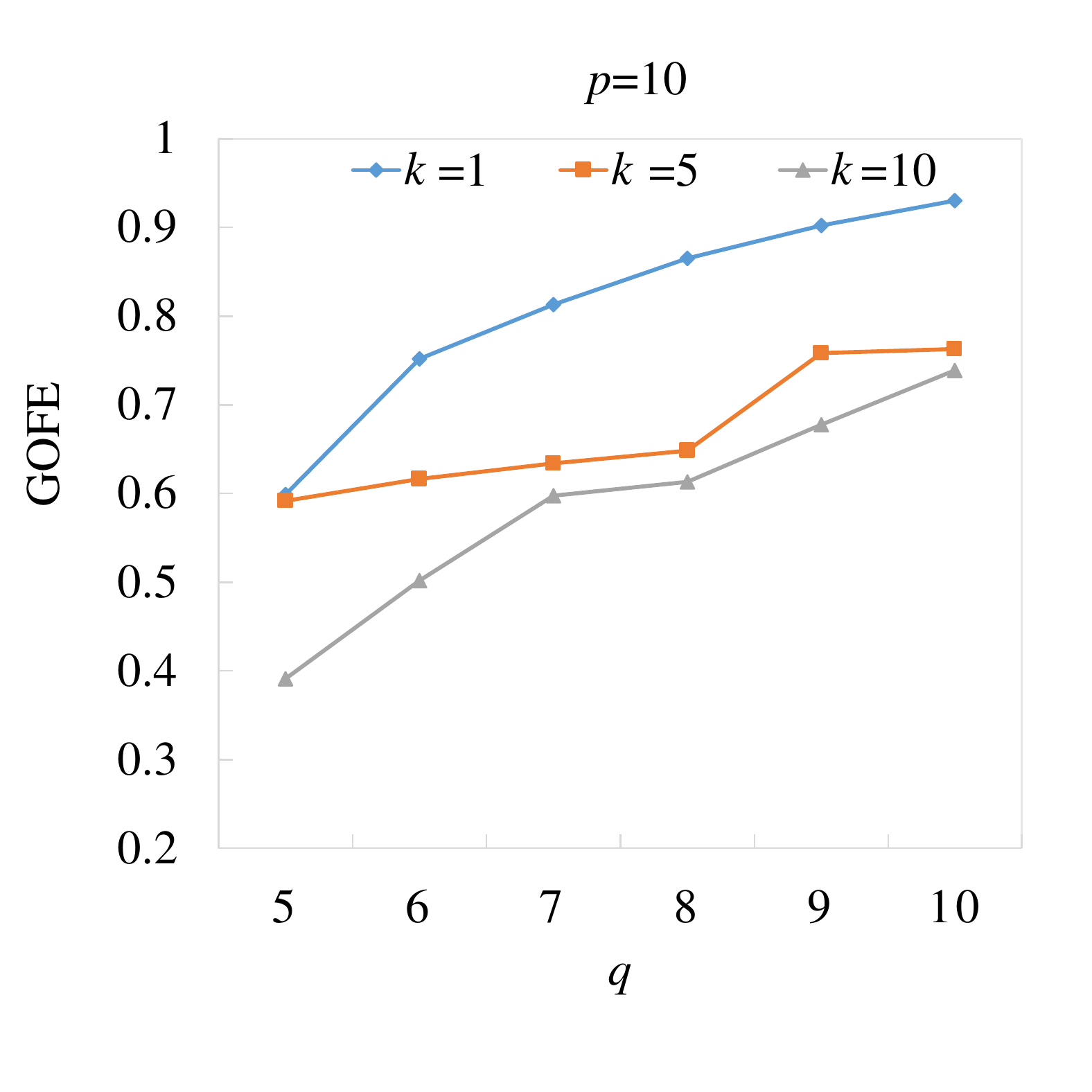}
	}
	\subfigure[Video Game.]{
		\includegraphics[width=0.25\linewidth]{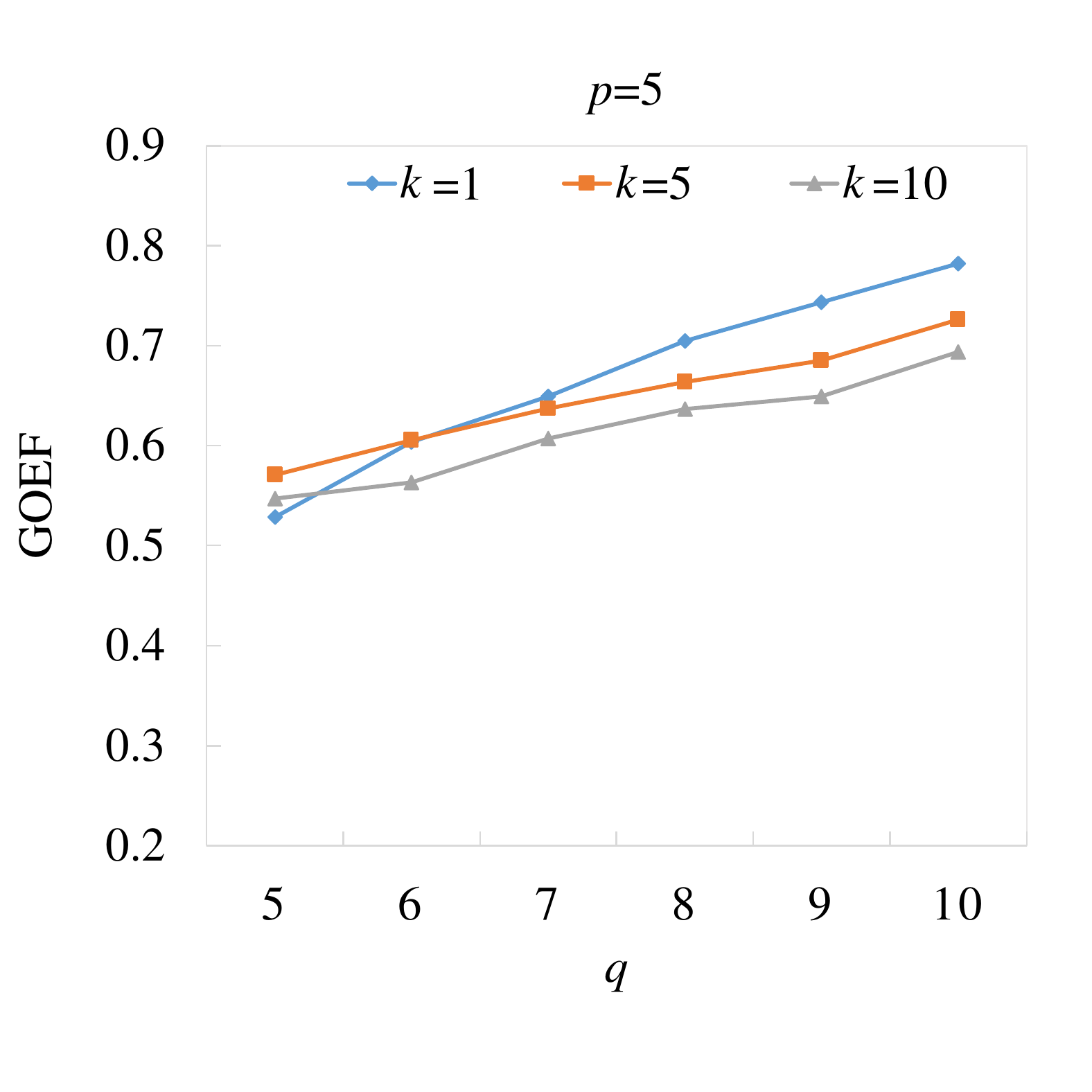}
	}
	\subfigure[Movie.]{
		\includegraphics[width=0.25\linewidth]{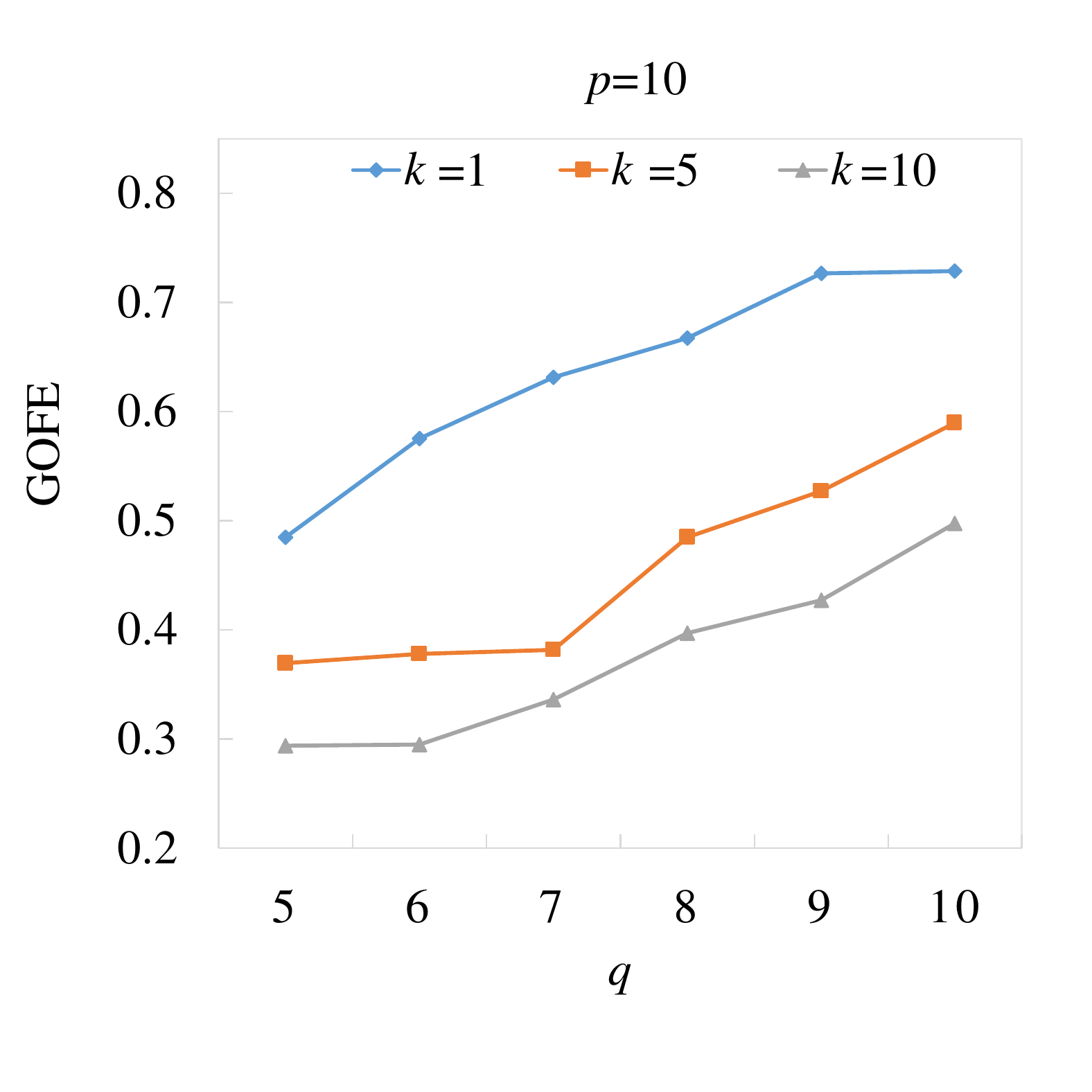}
	}
	\caption{GOFE at Fixed $p$.}
	\label{explain:compare}
\end{figure*}

\subsection{Rating prediction}
Table \ref{table:rmse} shows the rating prediction performance with respect to RMSE and MAE on the three datasets. First, we can see that the RMSE and MAE of HDE outperform the baseline methods on both datasets, which demonstrates the overall advantage of HDE due to its ability to generate the user/item embeddings with a fusion of two intermediate embeddings, the personalized embedding and temporal embedding. Particularly, due to the attentional network in the personalized embedding component, HDE can capture the different importance of items (users) to a user (an item) personalized preference, and due to the LSTM in temporal embedding component, HDE can capture the dynamic user aspect preference and item aspect quality.

We also note that the performance of HDE is better than that of NA-HDE and NL-HDE. Particularly, compared to NA-HDE, HDE reduces the RMSE by 3\%, 4\%, and 4\%, and reduces MAE by 5\%, 13\%, and 7\%, on Digital Music, Video Game, and Movie, respectively, which verifies the benefit brought by the personalized embedding and justifies our assumption that different items have different impact on the same user and different users have different impact on the same item. At the same time, compared to NL-HDE, HDE reduces the RMSE by 7\%, 12\%, and 6\%, and reduces MAE by 3.4\%, 10.7\%, and 12\%, on Digital Music, Video Game, and Movie, respectively. This result shows the effectiveness of the temporal embedding by which HDE can capture the dynamics of user aspect preference and item aspect quality from reviews.

\renewcommand\arraystretch{2}

\begin{table}[t]
	\caption{Performance of Rating Prediction.}
	\label{table:rmse}
	\centering
	\begin{tabular}{l |l l |l l |l l}
		\hline
		 &\multicolumn{2}{|c|}{Digital Music} &\multicolumn{2}{|c|}{Video Games}   &\multicolumn{2}{|c}{Movie}\\
		\cline{2-7}
		&\multicolumn{1}{|c}{RMSE} & MAE &\multicolumn{1}{|c}{RMSE} & MAE &\multicolumn{1}{|c}{RMSE} & MAE \\
		\hline
		PMF & 0.9418 & 0.6986 & 1.1119 & 0.8383 & 1.2606 & 0.9851\\
		HFT & 0.9184 & 0.6790 & 1.0709 & 0.7935 & 1.2247 & 0.9221\\
		EFM & 0.9072 & 0.6643 & 1.0935 & 0.8027 & 1.2331 & 0.9572\\
		DeepCoNN & 0.8875 & 0.6458 & 1.0620 & 0.7904 & 1.1311 & 0.8559\\
		NARRE & 0.8873 & 0.6541 & 1.0556 & 0.7922 & 1.1248 & 0.8196\\
		AMF & 0.8854 & 0.6370 & 1.0528 & 0.7527 & 1.0995 & 0.7766\\
		NA-HDE & 0.9031 & 0.6579 & 1.0975 & 0.8427 & 1.1217 & 0.8321\\
		NL-HDE & 0.9380 & 0.7092 & 1.0895 & 0.8260 & 1.1379 & 0.8749\\
		HDE & \textbf{0.8764} &  \textbf{0.6278} & \textbf{1.0526} & \textbf{0.7376} & \textbf{1.0742} & \textbf{0.7731}\\		
		\hline
	\end{tabular}
\end{table}

\begin{figure}[t]
	\centering
	\includegraphics[width=0.7\linewidth]{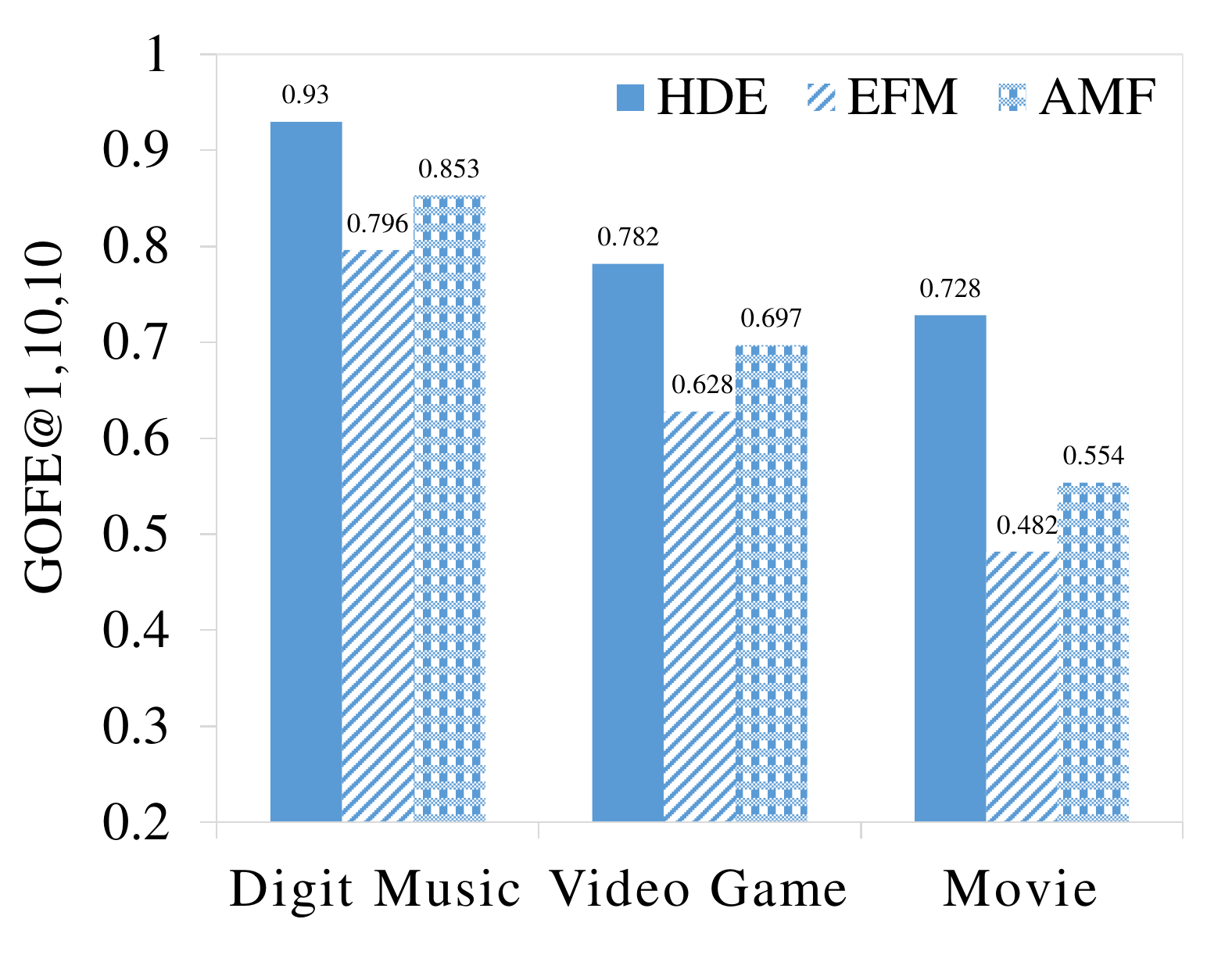}
	\caption{Comparison of Explainability.}
	\label{fig:GOFE_compare}
\end{figure}

\begin{figure*}[t]
\centering
	\subfigure[sample user profile.]{
		\includegraphics[width=0.20\linewidth]{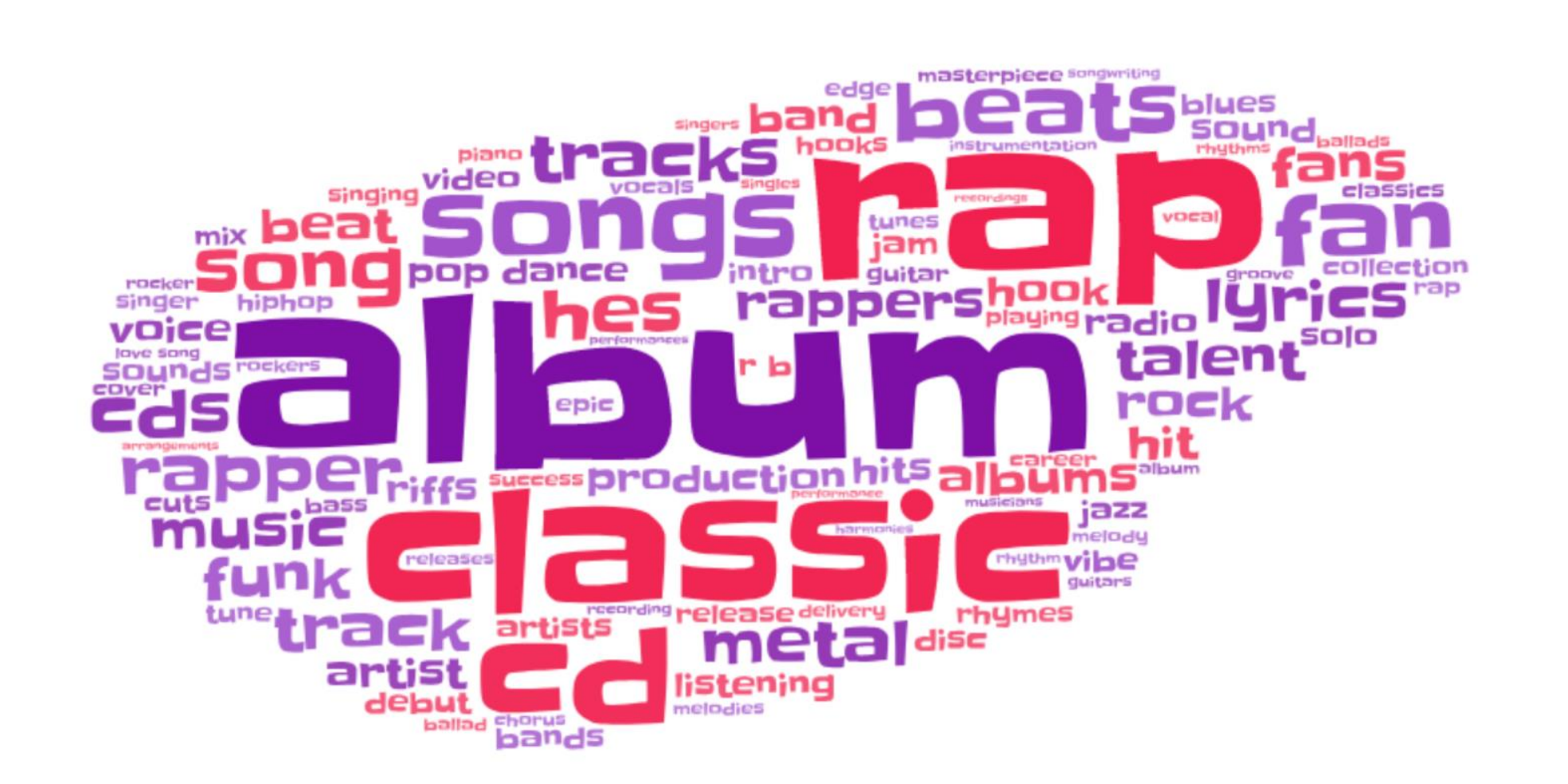}
	}
	\subfigure[item1 (recommended) profile.]{
		\includegraphics[width=0.20\linewidth]{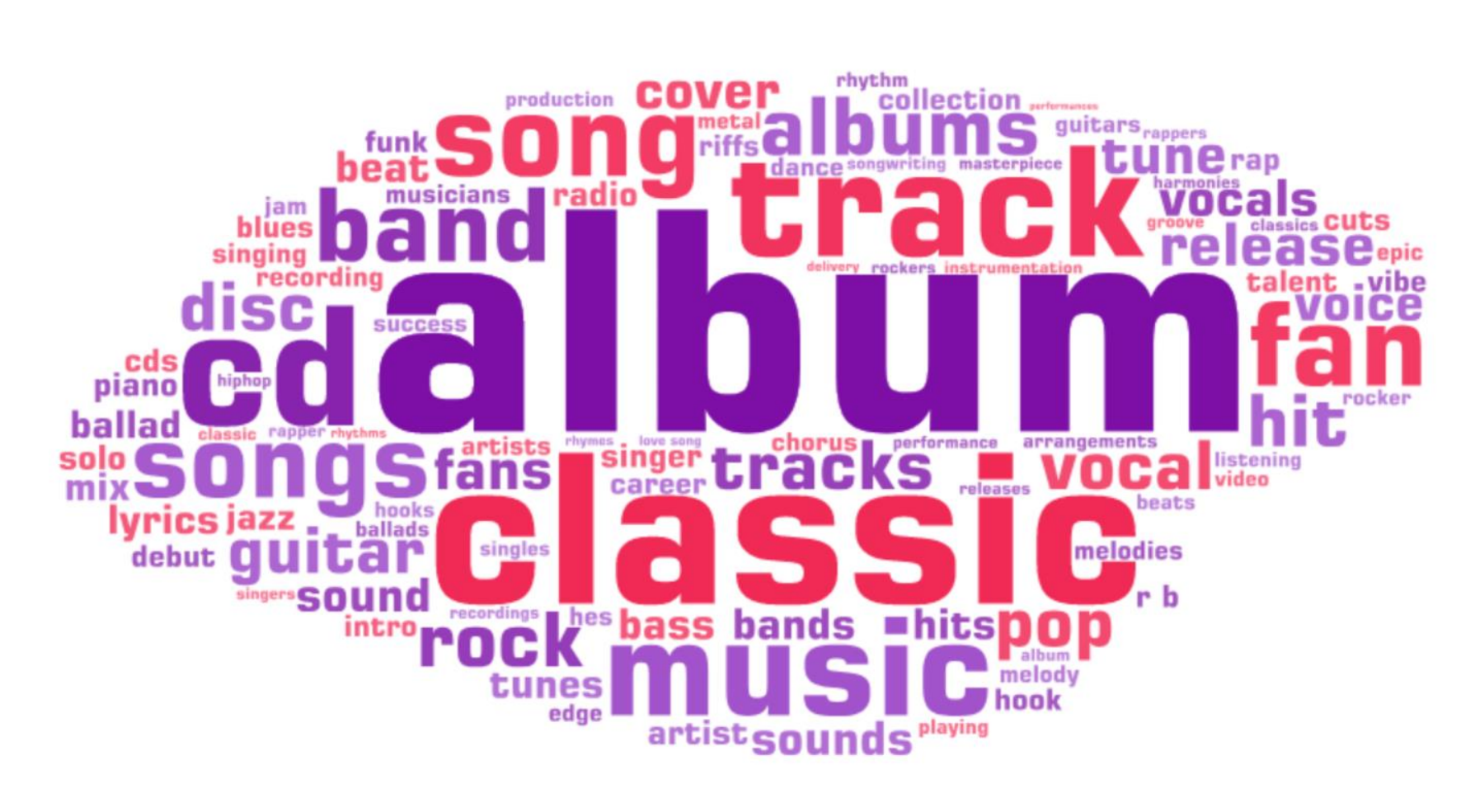}
	}
	\subfigure[item2 (recommended )profile.]{
		\includegraphics[width=0.20\linewidth]{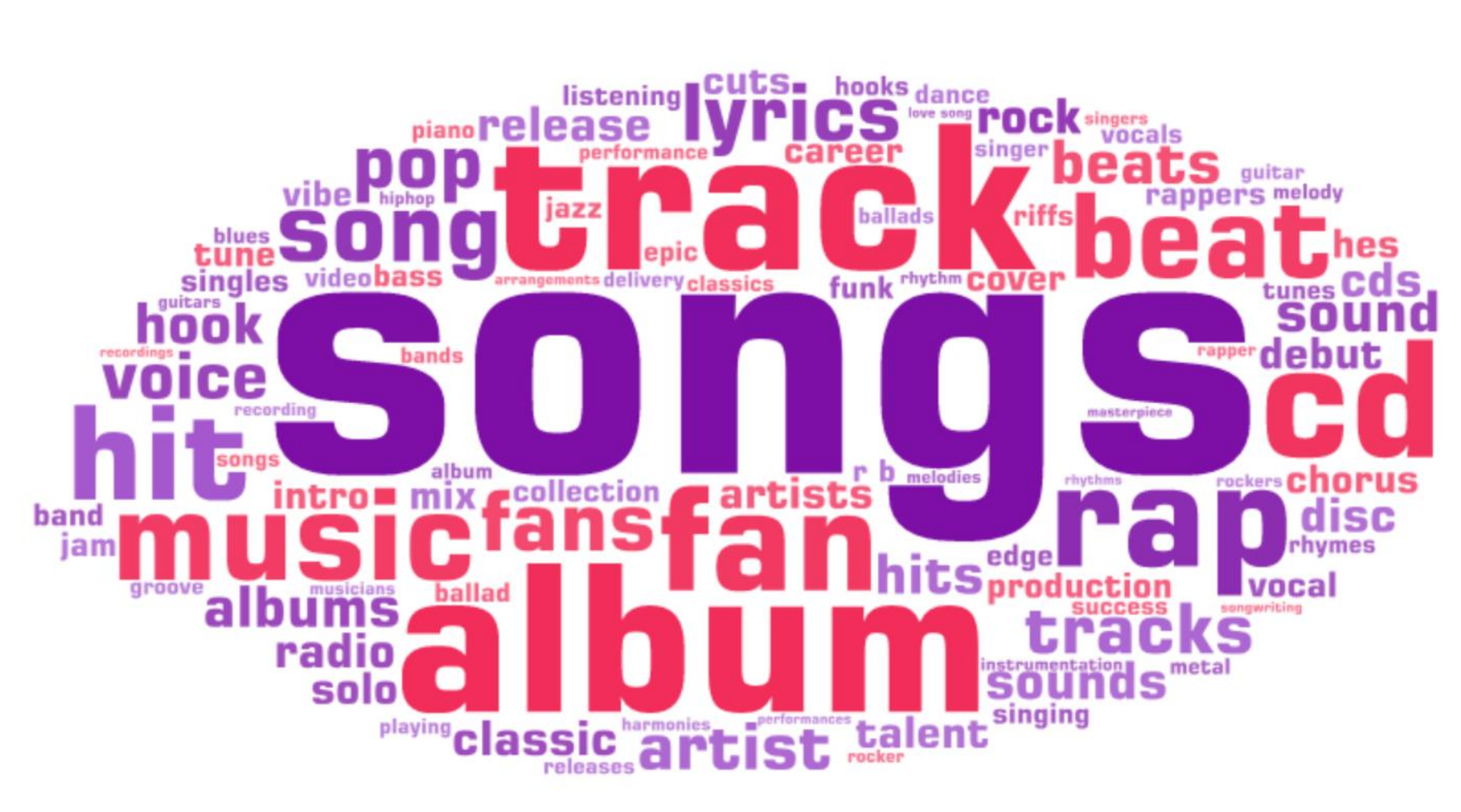}
	}	
	\subfigure[item3 (not recommended) profile.]{
		\includegraphics[width=0.20\linewidth]{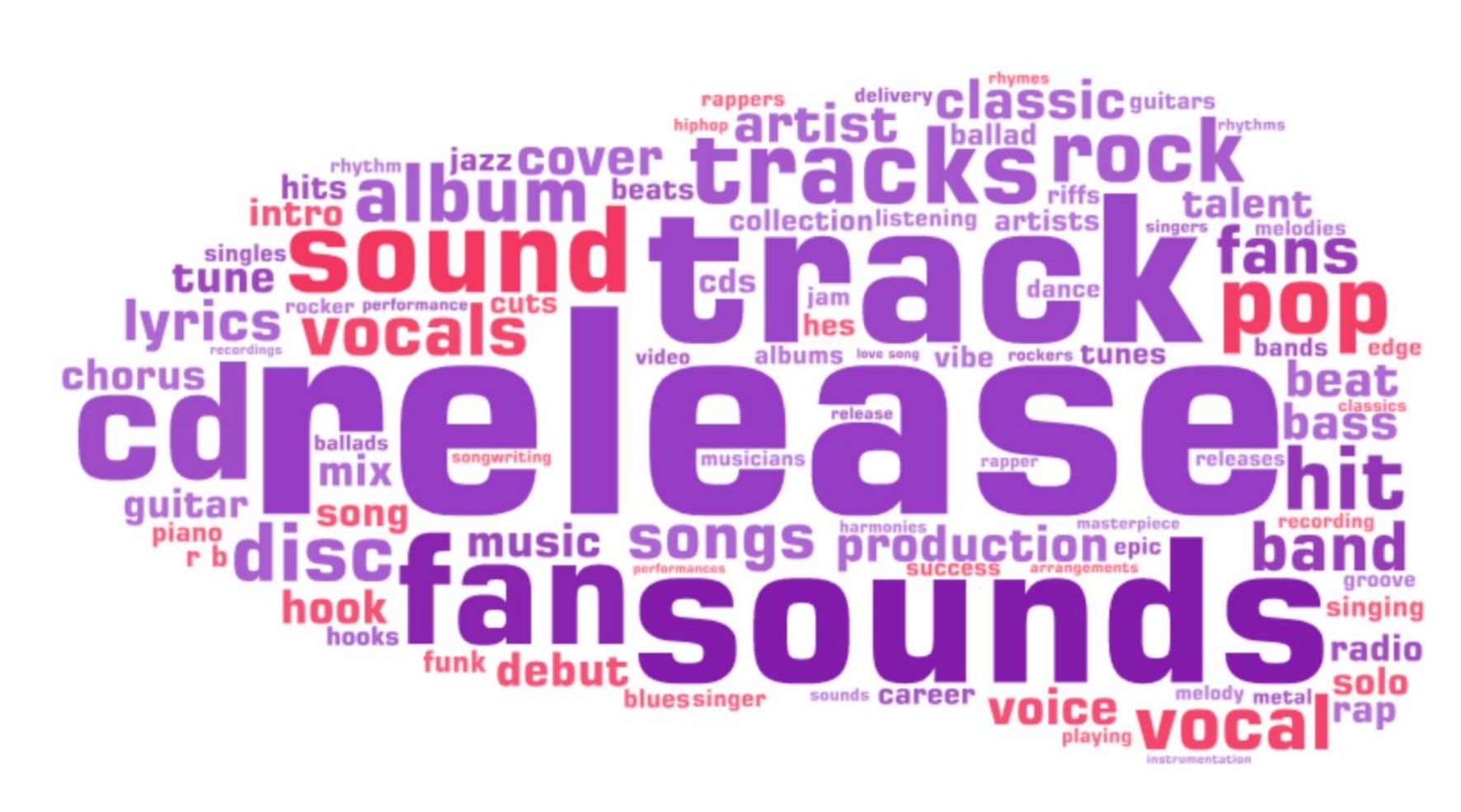}
	}
	
	\caption{The profiles of the sample user on Digit Music, two recommended items, and one not recommended item.}
	\label{fig:profiles}
\end{figure*}

\begin{figure*}[t]
\centering
	\subfigure[sample user profile.]{
		\includegraphics[width=0.20\linewidth]{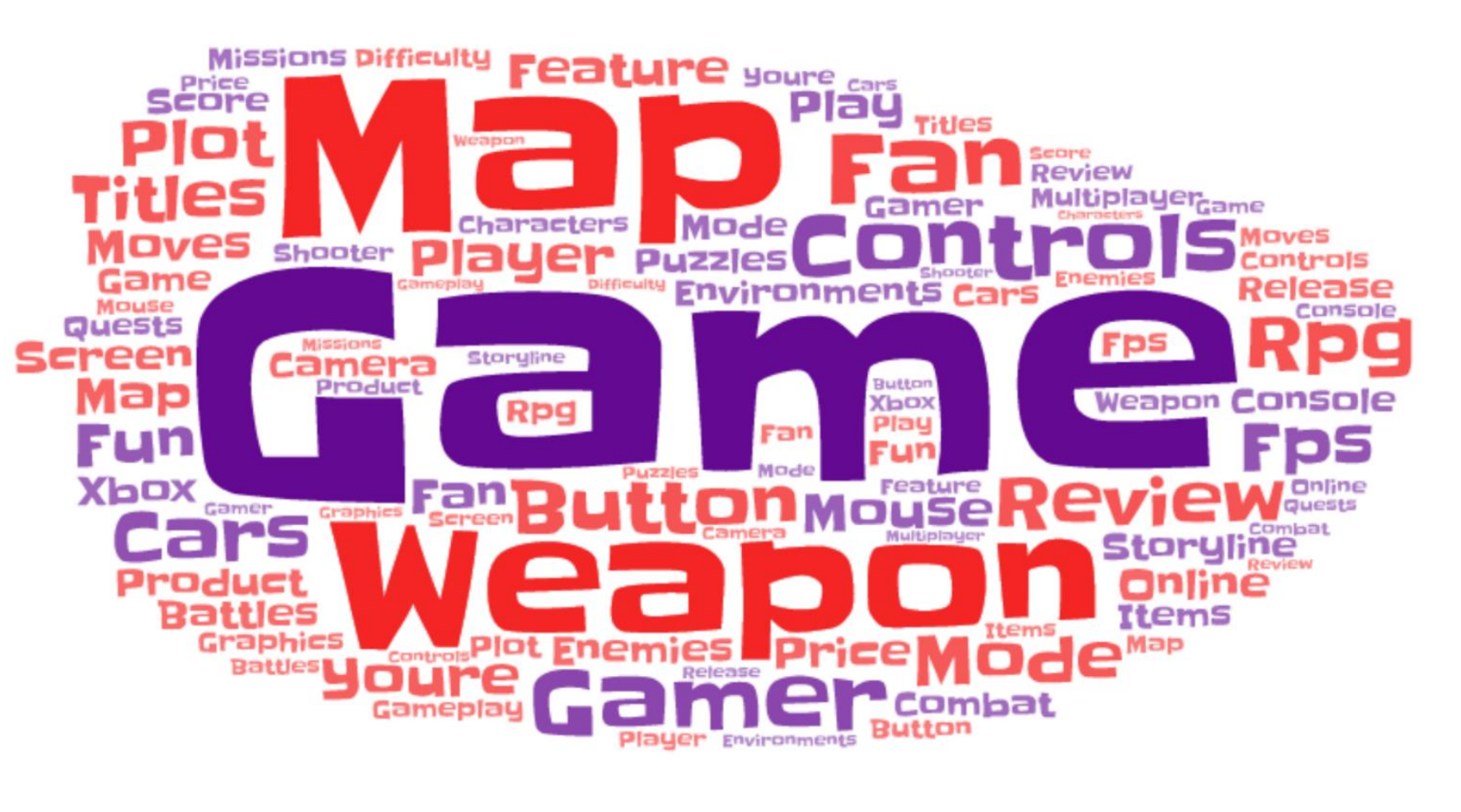}
	}
	\subfigure[item1 (recommended) profile.]{
		\includegraphics[width=0.20\linewidth]{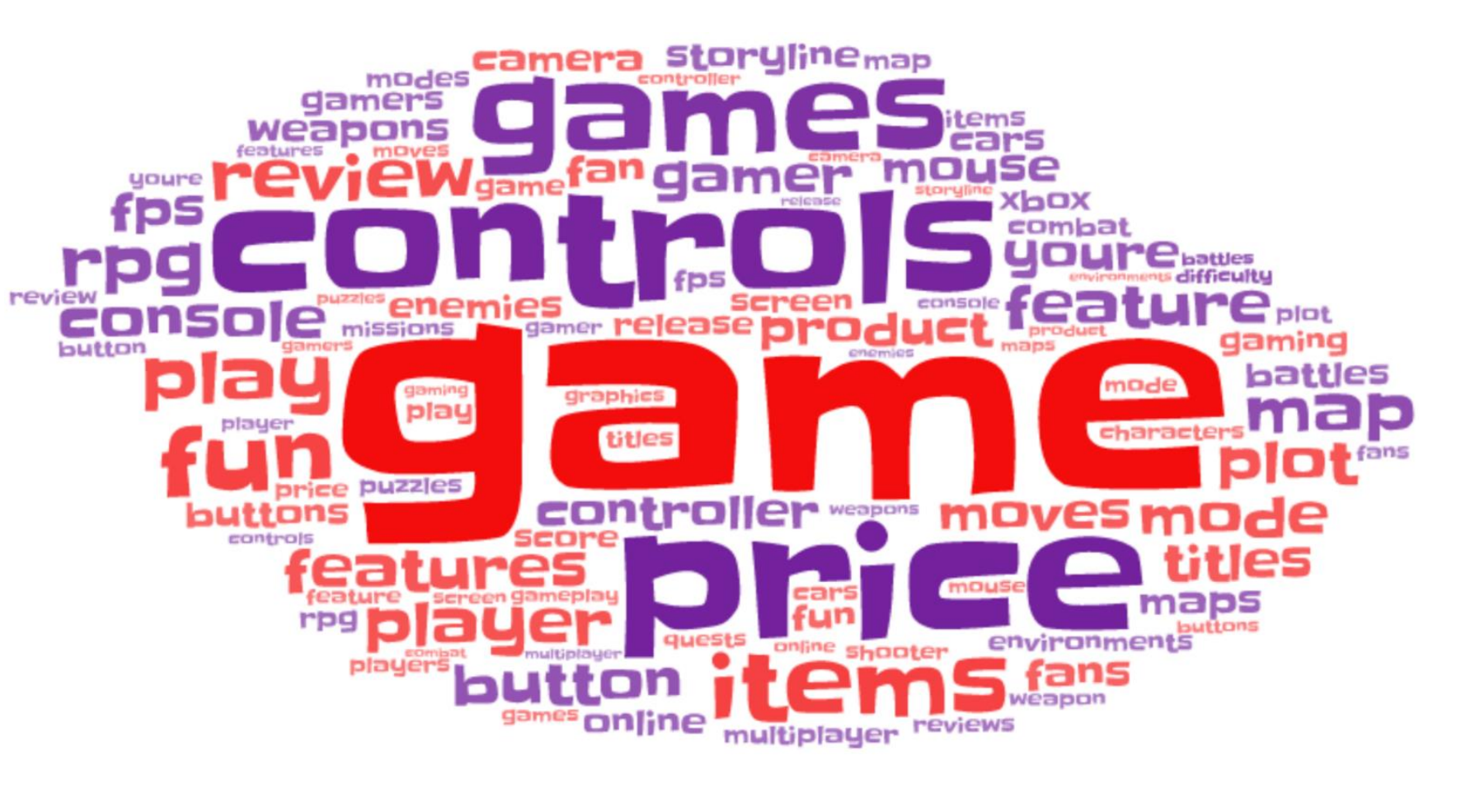}
	}
	\subfigure[item2 (recommended )profile.]{
		\includegraphics[width=0.20\linewidth]{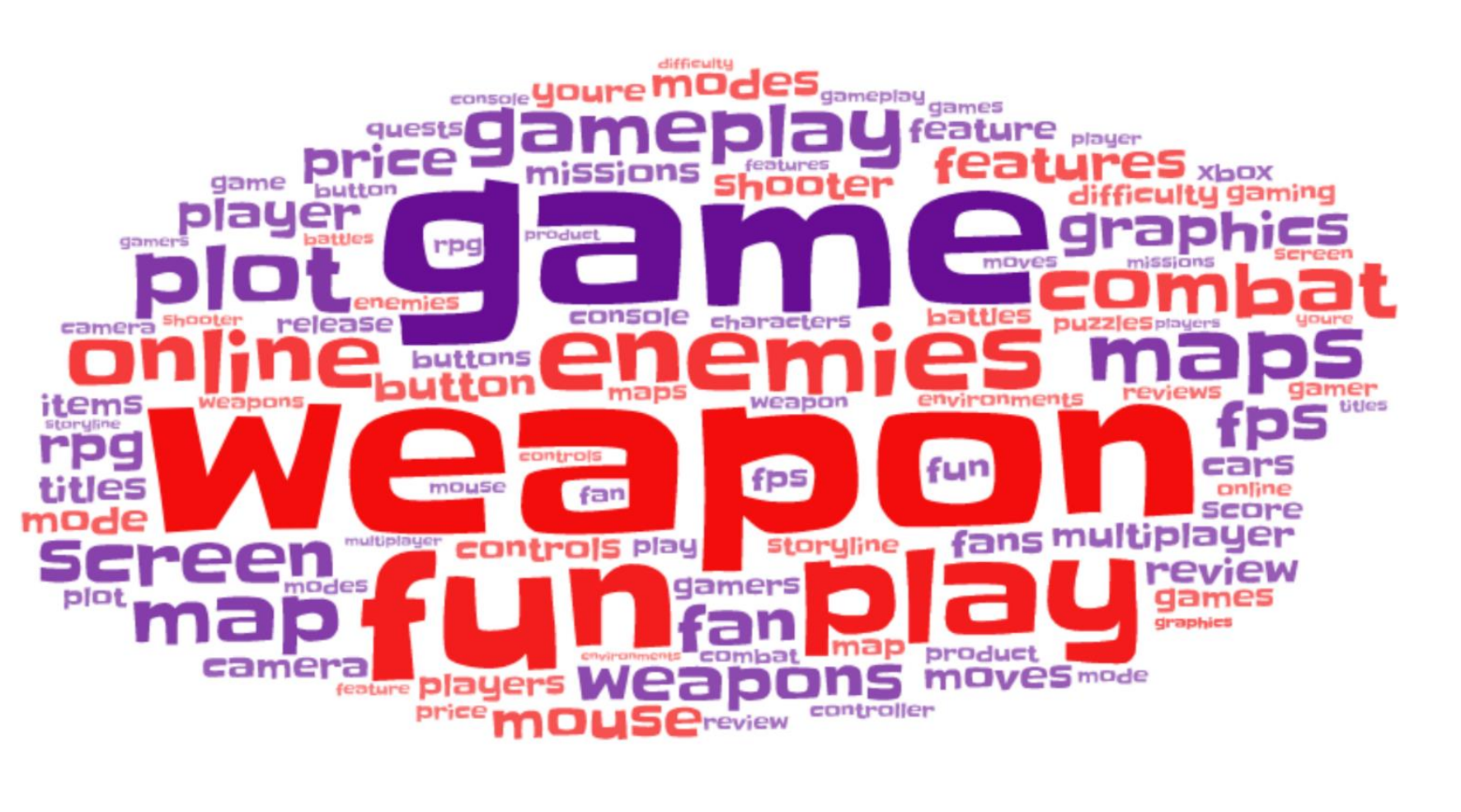}
	}	
	\subfigure[item3 (not recommended) profile.]{
		\includegraphics[width=0.20\linewidth]{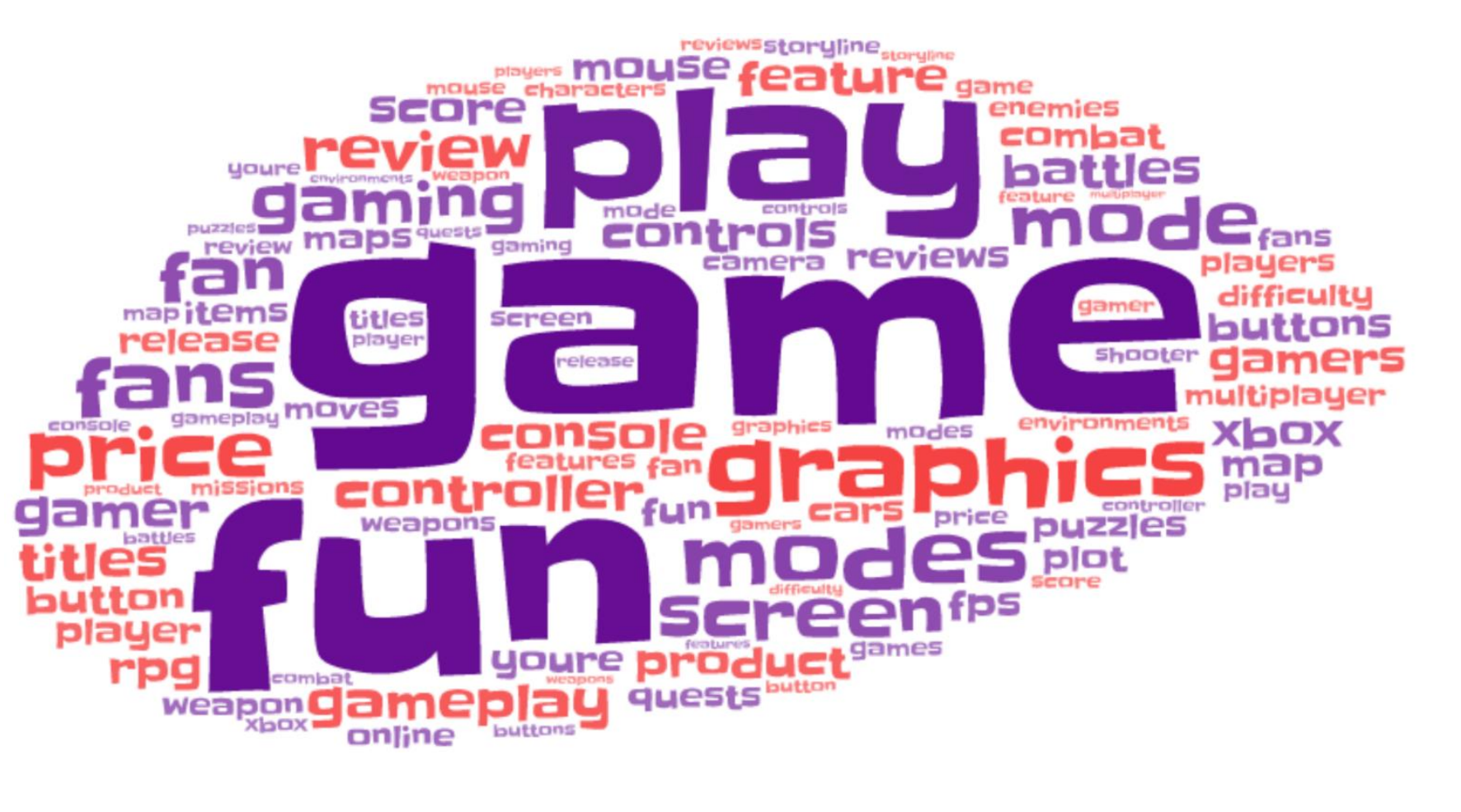}
	}
	
	\caption{The profiles of the sample user on Video Game, two recommended items, and one not recommended item.}
	\label{fig:profiles-vg}
\end{figure*}

\section{Evaluation of Explainability }

\subsection{Quantitative Evaluation of HDE Explainability}

Our idea to quantitatively evaluate the explainability of HDE is based on the intuition that the rationality of recommendations depends on wether the aspect quality of the recommended items satisfies the aspect preference of the user better than those of not recommended items.

For any aspect $a$, we sort all the items with respect to their quality on $a$, and choose top-$q$ items $\boldsymbol{Z}^{(a)}_q = \{ z_1, \cdots, z_q \} \subseteq \boldsymbol{V} $, i.e., for any $v \in Z^{(a)}_q$ and any $v' \notin Z^{(a)}_q$, $\hat{\boldsymbol{q}}^{(v)}(a) \ge \hat{\boldsymbol{q}}^{(v')}(a)$. At the same time, for a given user $u$, we first choose top-$p$ aspects $\boldsymbol{A}^{(u)}_p = \{ a_1, \cdots, a_p \}$ according to its aspect preference vector $\hat{\boldsymbol{p}}^{(u)}$, i.e., for any $a \in \boldsymbol{A}^{(u)}_p$ and any $a' \notin \boldsymbol{A}^{(u)}_p$ , $\hat{\boldsymbol{p}}^{(u)}(a) \ge \hat{\boldsymbol{p}}^{(u)}(a')$.

Suppose HDE recommends top-$k$ items $\boldsymbol{V}^{(u)}_k = \{v_1, $ $\cdots$$,$ $ v_k\}$ to user $u$ according to the predicted ratings. For any $v \in \boldsymbol{V}^{(u)}_k$, we need to check whether there is at least one aspect $a$ that is preferred by $u$, i.e., $a \in \boldsymbol{A}^{(u)}_p$, and whose quality $\hat{\boldsymbol{q}}^{(v)}(a)$ is better than that of the items not recommended, i.e., $v \in \boldsymbol{Z}^{(a)}_q$. For this purpose, we define the following identifier function:
\begin{equation}
\label{Eq_Identifier}
\mathbb{I}(u, v) = 
\begin{cases}
1, \ \exists a \in \boldsymbol{A}^{(u)}_p, v \in \boldsymbol{Z}^{(a)}_q \\
0,\ \text{otherwise}.
\end{cases}
\end{equation}
Basically, $\mathbb{I}(u, v) = 1$ implies that the explanation why $v$ is recommended to $u$ is that item $v$ is satisfied by user $u$ due to some aspect $a$ preferred by $u$ on which $v$ is better than other items. 
Now we can define the following metric called Goodness Of Fit on Explanation (GOFE) to measure the explainability of HDE,
\begin{equation}
GOFE@k,p,q = \frac{\sum_{u \in \boldsymbol{U}}c^{(u)}}{C},
\end{equation}
where $c^{(u)} = \sum_{v\in \boldsymbol{V}^{(u)}_k} \mathbb{I}(u, v)$ is the number of recommended items satisfied by $u$, and $C = |\boldsymbol{U}|*k$ is the total number of items recommended to all the users. Essentially, GOFE can be understood as the probability that HDE can give explanations from the perspective of preference satisfaction. 

Figure \ref{explain:K=1} shows the GOFE at $k = 1$ on the three datasets, which means we only recommend one item to users. We can see that GOFE increases with $q$ and $p$ on all the three datasets. Basically, $q$ and $p$ define the scope of candidate explanations from the perspective of item and the perspective of aspect, respectively. The results, therefore, are consistent with the intuition that larger the scope of possible explanations, better the explainability. 

Figure \ref{explain:compare} shows the GOFE at fixed $p$ on the three datasets. As the aspects of Digit Music and Movie are more than those of Video Game, we set $p=10$ on Digit Music and Movie while $p=5$ on Video Game. We can see that GOFE increases with $q$, again due to more candidate explanations incurred by larger $q$. We can also note that GOFE increases with $k$, which indicates an interesting property of HDE that the more recommended items, greater the explainability of HDE.

To further verify the explainability fo HDE, we also compare it with EFM and AMF, which are most similar to our work as they can also provide aspect-level explanations. To make the comparison fair, we set $p$, $q=10$ at which EFM and AMF perform best. As we can see from Figure \ref{fig:GOFE_compare}, the GOFE of HDE significantly outperforms that of EFM and AMF, which indicates that HDE has better explainability than EFM and AMF. We argue that this is because of two reasons. One is that the dynamic aspect-level explanations offered by HDE are more proper than the static ones offered by EFM and AMF. The other reason is that HDE can capture the preference of a user to aspects even if they are not mentioned by that user.

\subsection{Case Study for Explainability Verification}


At first, on Digit Music, we first randomly sample one user with ID "mistermaxxx08", and then visualize her/his aspect preference vector $\hat{\boldsymbol{p}}$ learned by HDE using a word cloud shown in Figure \ref{fig:profiles}(a), where each word represents an aspect and the size of the word representing aspect $i$ is proportional to the $\hat{\boldsymbol{p}}(i)$ which indicates how much the preference of the user to that aspect. From Figure \ref{fig:profiles}(a) we can see the top-5 aspects preferred by the sample user are "\textbf{album}, \textbf{classic}, \textbf{rap}, \textbf{cd}, \textbf{songs}". If we expand the range to top-$10$, we can see the aspects "beats" and "hes", which are not mentioned in the reviews of the user, are included. Such result shows that HDE is able to learn user latent preference to aspects even though they are not mentioned in user reviews. 

The top-2 items recommended by HDE to this user are item1 (with ID B0000004UM) and item2 (ID B0000004YB), whose aspect quality vectors are shown in \ref{fig:profiles}(b) and \ref{fig:profiles}(c), respectively. At the same time, we also randomly choose one item (item3 with ID B0009VJWQS) not recommended and show its aspect quality vector in \ref{fig:profiles}(d). From Figures \ref{fig:profiles}(b), \ref{fig:profiles}(c), and \ref{fig:profiles}(d), we can see that item1 performs well on the aspects "\textbf{album}, \textbf{classic}, \textbf{cd}, track, band", item2 performs well on "\textbf{songs}, \textbf{album}, \textbf{cd}, \textbf{rap}, fan", and item3 performs well on "release, sounds, track, \textbf{cd}, pop". It is obvious that the aspect quality of item1 and item2 is more consistent with the user aspect preference than item3 is. Particularly, for the recommendation of item1, we can generate the explanation as "You might be interested in [album, classic, cd], on which item1 performs well", while for the recommendation of item2, we can generate the explanation as "You might be interested in [album, rap, cd], on which item2 performs well". 

Similarly, we also sample one user on Video Game and use HDE to recommend top-2 items to her/him, whose aspect preference/quality vectors are visualized in Figure \ref{fig:profiles-vg}. Again, we can see that the aspect quality of recommended items (shown in Figures \ref{fig:profiles-vg}(b) and \ref{fig:profiles-vg}(c)) are more consistent with the aspect preference of the sample user (shown in \ref{fig:profiles-vg}(a)) than that of not recommended item (shown in \ref{fig:profiles-vg}(d)).

\subsection{Case Study for Capturing Dynamic Preference}

As we have mentioned before, preference of users always change over time. Again, we use examples to show the ability of HDE to capture the user dynamic preference. For the sample users  same as above, HDE generates their aspect preference vectors at 2008 and 2004, which are visualized in Figures \ref{fig:user-profiles}(a) and Figure \ref{fig:user-profiles}(b) for the sample user on Digit Music, and Figures \ref{fig:user-profiles-vg}(a) and Figure \ref{fig:user-profiles-vg}(b) for the sample user on Video Game, respectively. From Figures \ref{fig:user-profiles} and \ref{fig:user-profiles-vg}, we can see that in 2014, these users had new preferences which they did not have in 2008 . For example, in 2014 the sample user on Digit Music became interested in classic music which was not her/his preference in 2008.

\begin{figure}[t]
\centering
	\subfigure[sample user profile at 2008.]{
		\includegraphics[width=0.40\linewidth]{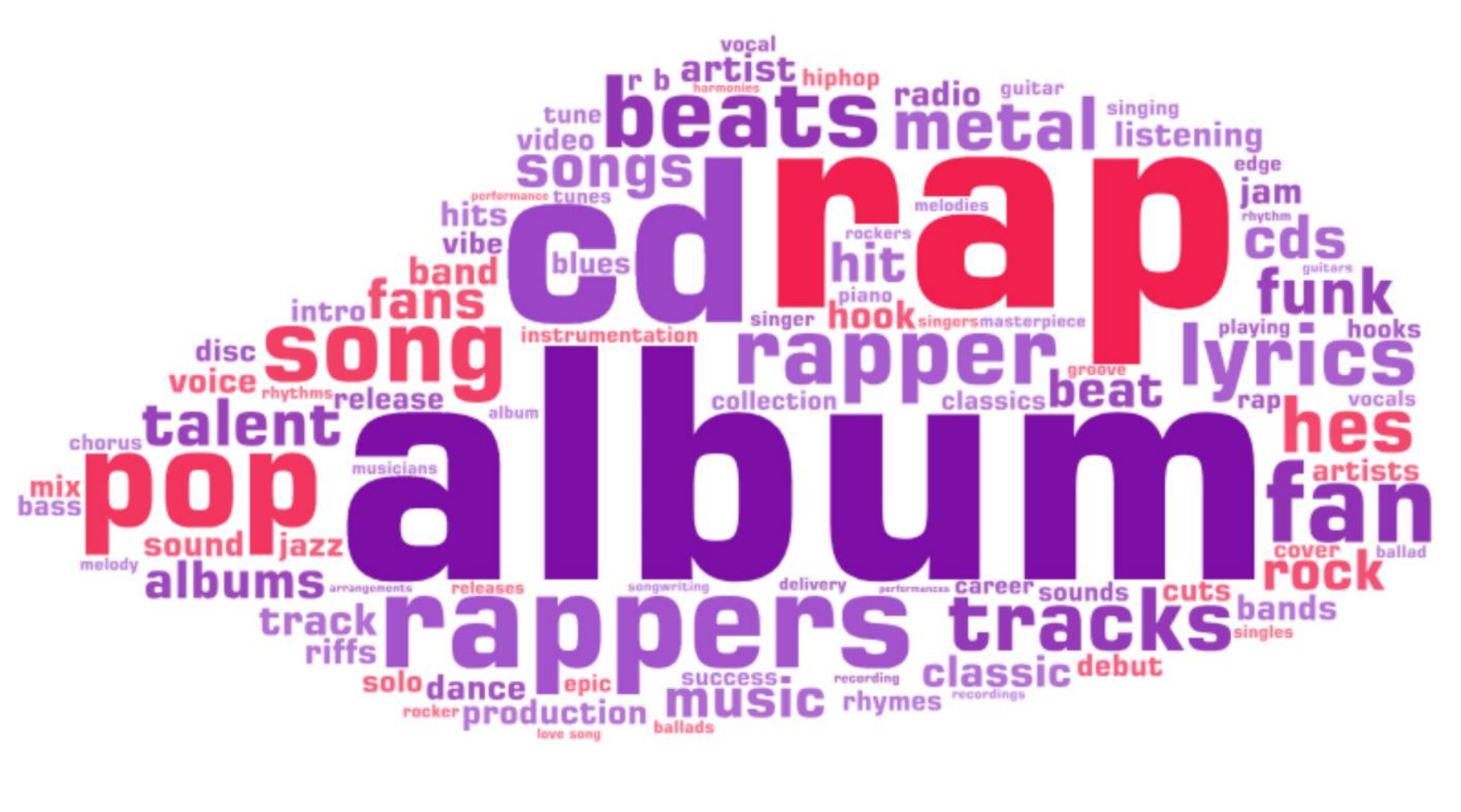}
	}
	\subfigure[sample user profile at 2014.]{
		\includegraphics[width=0.40\linewidth]{profile.pdf}
	}
	\caption{The dynamic profiles of the sample user on Digit Music.}
	\label{fig:user-profiles}
\end{figure}

\begin{figure}[t]
\centering
	\subfigure[sample user profile at 2008.]{
		\includegraphics[width=0.40\linewidth]{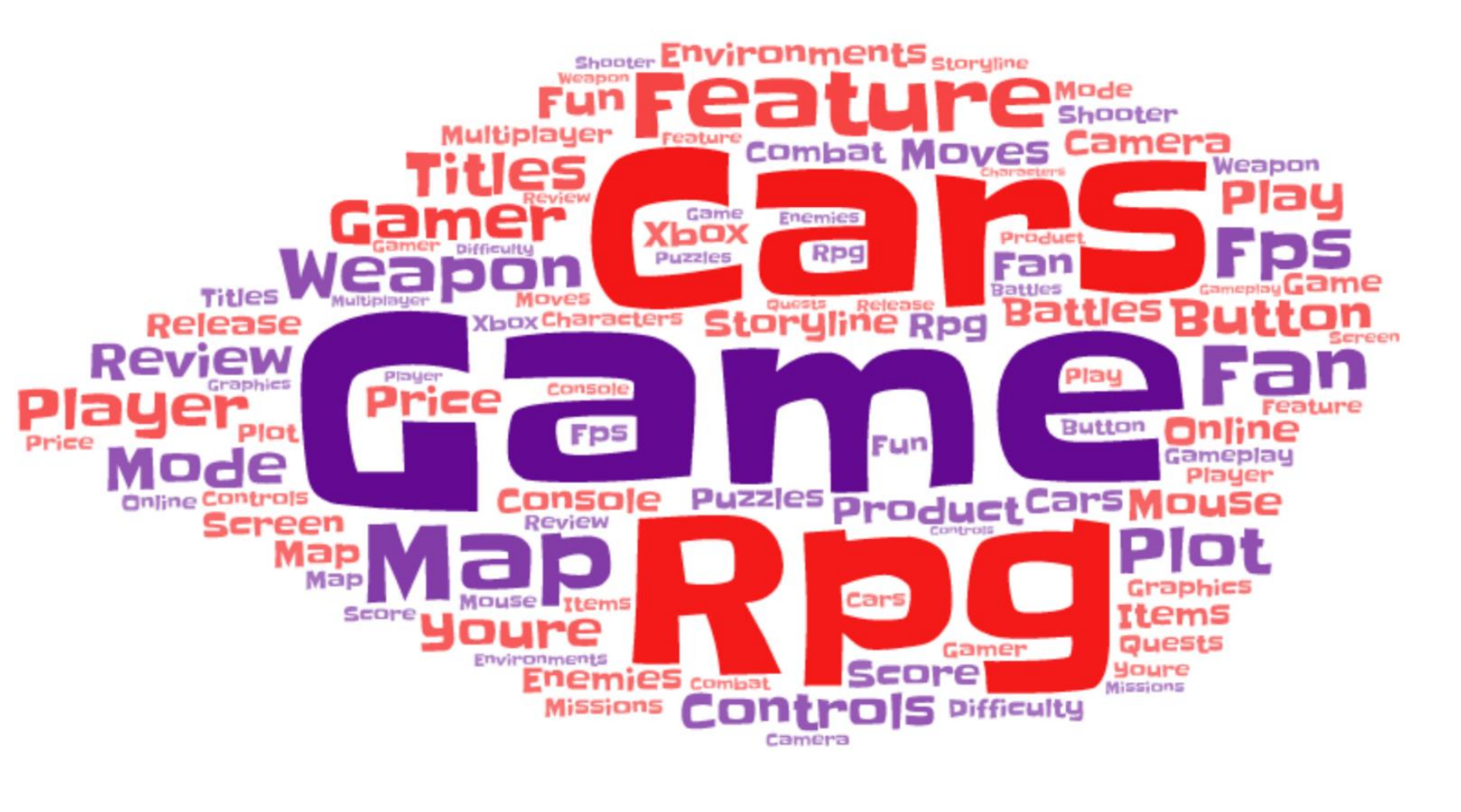}
	}
	\subfigure[sample user profile at 2014.]{
		\includegraphics[width=0.40\linewidth]{user-profile-vg-2014.pdf}
	}
	\caption{The dynamic profiles of the sample user on Video Game.}
	\label{fig:user-profiles-vg}
\end{figure}

\section{RELATED WORK}\label{section:related_work}

%

\subsection{Explainable Recommendation}
The existing methods for explainable recommendation roughly fall into two classes. One class of the explainable recommendation methods generate explanations based on relevant users or items, where a recommendation of an item can be explained as "the users who are similar to you like the item", or "the item is similar to the items you like" \cite{b19,b20}. The other class is based on based on reviews. Recently, a large number of literatures have been proposed for exploiting textual review information to provide explanations while improving the rating prediction performance, for examples, EFM\cite{b14}, HFT\cite{b6}, AMF\cite{b3}, and NARRE\cite{b1}. 

Recently, a large number of literatures have been proposed for exploiting textual review information to provide explanations while improving the rating prediction performance, for examples, EFM\cite{b14}, HFT\cite{b6}, AMF\cite{b3}, and NARRE\cite{b1}. Aspect-based explainable recommendation methods extract aspect information from review, where two types of aspects are defined, one is defined as a noun word or phrase that represents a feature \cite{b14}, and the other is defined as a set of words that describe a topic in the reviews \cite{b3,b30,cheng2018aspect}.  Zhang et al. propose a model that extracts explicit product features and user opinions by phrase-level sentiment analysis, and then uses Matrix Factorization to produce the recommendation \cite{b14}. Hou et al. propose an Aspect-based Matrix Factorization (AMF) model which can make recommendations by fusing auxiliary topic-based aspect information extracted from reviews into matrix factorization \cite{b3}. McAuley et al. propose an approach that combines latent rating dimensions with latent review topics, which uses an exponential transformation function to link the topic distribution over reviews \cite{b6}.  Li et al. propose a deep learning based framework named NRT which leverages gated recurrent units (GRU) to summarize the massive reviews of an item and generate tips for an item \cite{b31}. Recently, some works that provide review-level explanations have been also proposed. For example, Chen et al. propose an attention mechanism based model to explore the usefulness of reviews and produce highly-useful review-level explanations to help users make decisions \cite{b1}.

\subsection{Deep Learning for Recommendation}


Recently, some research works have incorporated deep learning techniques, including RBM \cite{deng2016deep}, Autoencoders \cite{wei2017collaborative}, RNN \cite{Xu:2019}, and CNN \cite{wu2017content}, into recommender systems to improve the performance of user and item embedding learning. In addition to combining deep neural networks with collaborative filtering \cite{b1}, the existing deep learning based recommendation models often integrate textual reviews to enhance the performance of latent factor modeling \cite{b28,b10,b14,b3,b29}. For example, DeepCoNN\cite{b10} uses convolutional neural networks to process reviews, and utilizes deep learning technology to jointly model user and item from textual reviews. Recently, some works have incorporated attention mechanism into recommender systems\cite{b18,b22,wang2018attention,kang2018self,ma2019gated,cheng20183ncf}. However, the existing works based on deep learning often only focus on the latent feature learning for users and items, but ignore the explainability of recommendations.

\section{Conclusions}\label{section:conclusions}

In this paper, we propose a novel model called Hybrid Deep Embedding (HDE) for recommendations with dynamic aspect-level explanations. We introduce a hybrid embedding framework by which HDE can make recommendations by fusing dynamic aspect information extracted from reviews with user-item interactions. HDE first learns two intermediate embeddings, Personalized Embedding (IE) and Temporal Embedding (TE) for capturing the dynamic personalized preference, and then generate the finally embeddings of users and items for rating prediction. Simultaneously, HDE can generate the dynamic aspect preference/quality vectors for users/items via an encoder-decoder based network. The results of the extensive experiments conducted on real datasets verify the recommendation performance and explainability of HDE.

\section*{Acknowledgment}
This work is supported by National Natural Science Foundation of China under grant 61972270, Hightech Program of Sichuan Province under grant 2019YFG0213, and in part by NSF under grants III-1526499, III-1763325, III-1909323, CNS-1930941, and CNS-1626432  

\bibliographystyle{abbrv}
\bibliography{HDE} 

\balance

\end{document}